\documentclass[12pt]{article}
\usepackage{graphics,graphicx}
\usepackage{amssymb,epsfig,amsmath,euscript,array}
\usepackage[nosort]{cite}
\usepackage{axodraw4j}
\usepackage{pstricks}
\usepackage{color}
\usepackage{simplewick}

\def\a{\alpha}

\def\g{\gamma}

\def\s{\sigma}

\newcommand{\cE}{\mathcal E}

\newcommand{\cM}{\mathcal M}
\newcommand{\cN}{\mathcal N}
\newcommand{\cO}{\mathcal O}

\newcommand{\cZ}{\mathcal Z}

\newcommand{\be}{\begin{equation}}
\newcommand{\bea}{\begin{eqnarray}}

\newcommand{\ee}{\end{equation}}
\newcommand{\eea}{\end{eqnarray}}

\def\g{\gamma}

\def\mP{ \mathbb{P}}

\def\mZ{ {\mathbb{Z}}  }

\def\Aut{ {\rm{ Aut }} }

\begin{document}

\makeatletter
\@addtoreset{equation}{section}
\makeatother
\renewcommand{\theequation}{\thesection.\arabic{equation}}

\rightline{QMUL-PH-12-13}
\rightline{WITS-CTP-104}
\vspace{1.8truecm}

\vspace{15pt}

%%%%%%%%%%%%%%%%%

{\LARGE{ 
\centerline{ \bf  On the refined counting of graphs on surfaces  } 
}}  

\vskip.5cm 

\thispagestyle{empty} \centerline{
    {\large \bf Robert de Mello Koch
${}^{a,} $\footnote{ {\tt robert@neo.phys.wits.ac.za}}}
   {\large \bf  Sanjaye Ramgoolam ${}^{b,}$
\footnote{ {\tt s.ramgoolam@qmul.ac.uk}} , Congkao Wen
               ${}^{b,}$ \footnote{\tt c.wen@qmul.ac.uk}  }
                                                       }

\vspace{.4cm}
\centerline{{\it ${}^a$ National Institute for Theoretical Physics ,}}
\centerline{{\it Department of Physics and Centre for Theoretical Physics }}
\centerline{{\it University of Witwatersrand, Wits, 2050, } }
\centerline{{\it South Africa } }

\vspace{.4cm}
\centerline{{\it ${}^b$ Centre for Research in String Theory, Department of Physics},}
\centerline{{ \it Queen Mary University of London},} \centerline{{\it
    Mile End Road, London E1 4NS, UK}}

\vspace{1.4truecm}

%%%%%%%%%%%%%%%%%
\thispagestyle{empty}

\centerline{\bf ABSTRACT}

\vskip.4cm 

Ribbon graphs embedded on a Riemann surface 
provide a useful way to describe the double line Feynman 
diagrams of large $N$ computations and a variety of other 
QFT correlator and scattering amplitude calculations, e.g in MHV rules for scattering amplitudes, 
as well as in ordinary QED. Their counting is a special case of the 
counting of bi-partite embedded graphs. 
We review and extend relevant mathematical literature and 
 present results on the counting of some infinite classes of 
bi-partite graphs. Permutation groups and representations 
as well as double cosets and quotients of graphs are useful mathematical tools. 
The counting results are refined according to data 
of physical relevance, such as the structure of the vertices, faces and genus 
of the embedded graph. These counting problems can be expressed  
in terms of observables in three-dimensional topological field theory
with $S_d$ gauge group which gives them  a topological membrane interpretation.

\setcounter{page}{0}
\setcounter{tocdepth}{2}

\newpage

\tableofcontents

\setcounter{footnote}{0}

\linespread{1.1}
\parskip 4pt

{}~
{}~

\newpage

\section{ Introduction }

Graphs embedded on surfaces have a variety of physical applications
in the context of large $N$ expansions of quantum field theories (QFTs). 
The Feynman rules of large $N$ gauge theories can be organized by these 
embedded graphs with the leading contribution coming from 
the spherical topology and subleading contributions from higher genus
surfaces \cite{'tHooft}.  The computation of a correlator or amplitude in 
4 or general dimensions combines the combinatoric element of 
enumerating these graphs, computing the associated integrals and summing 
over the graphs. In matrix models, which can be viewed as QFTs in zero 
dimensions, correlators are reduced to the combinatoric elements of 
 perturbative QFT problems. In  hermitian matrix models, the correlators can be expressed 
in a very simple form in terms of triples of permutations \cite{MR1}. 
These can be related to the usual double-line diagram method of 
large $N$ calculations, which can be equivalently thought as ribbon graphs, 
i.e.  graphs with vertices and edges, but with a cyclic order at the 
vertices, which can be viewed as inherited from the orientation of the 
embedding surface. The link between ribbon graphs and 
the permutation triples becomes 
easiest to see when we divide the edges of the ribbon graphs by introducing 
a new type of vertex, let's call them white vertices, while
the previously present vertices are called black vertices.
These form an example of embedded bi-partite graphs, where the edges 
only link black to white. The bi-partite graphs just described can be 
generalized to allow both black and white vertices to have general valencies. 
 These more general bi-partite graphs show up in calculations of correlators 
in  complex matrix models of relevance to the half-BPS sector of 
maximally supersymmetric Yang-Mills theory (SYM) \cite{cjr,tom}.

The counting of ribbon graphs was studied in \cite{MR2},
alongside other Feynman graph counting problems,
using a uniform method for dealing with ordinary (non-large-$N$)
problems and ribbon graph problems. Generating functions for 
various graph counting problems of physical interest 
were given using cycle indices, which keep track of the cycle structures 
of the permutations in a subgroup of the symmetric group $S_d$ - the group 
of all the $d!$ permutations of a set of $d$ elements, usually 
taken as $\{ 1, \cdots , d \}$.  Following the approach of \cite{Read}, 
the key mathematical tools  were the Burnside Lemma from combinatorics 
(described briefly in the Glossary)  and double cosets of permutation groups.  The formulae which lead to the simplest 
counting were found to have a very nice interpretation in terms 
of topological field theory with $S_d$ gauge group on a cylinder or on a torus. 
This could be interpreted in terms of counting problems for strings 
mapping to a two dimensional target, using similar logic to that which
 lead to the discovery of the string theory dual to large $N$ two-dimensional 
Yang Mills theory \cite{grta}. The 3-holed sphere, the cylinder as well 
as the torus played a role. Much of the story held true for  Feynman 
graphs without large $N$,  such as ordinary scalar field theory
 or Quantum Electrodynamics (QED). Along the way, a somewhat surprising 
result was that there is a map between the counting of QED vacuum graphs 
and ribbon graphs.

In section 2, we give a review of some of the background 
material which serves as motivation for the counting problems 
we consider in this paper. Aside from matrix models and the combinatorics 
of large $N$ correlators, ribbon graphs also arise in the context of 
MHV rules for constructing amplitudes \cite{CSW,BST}. We explain 
how the permutation triple description of ribbon graphs 
can be used to extract the trace structure of amplitudes.

This paper extends the permutation group based counting of  
embedded graphs to a refined counting where we fix the structure of the 
vertices as well as the faces. The permutation group approach continues 
to give surprising insights on the geometrical meaning of 
the counting problems revealing, 
 in this refined case,  the role of topological field theory 
with $S_d$ gauge group in three dimensions, which in turn can 
be interpreted in terms of a topological membrane counting.

For this refined counting problem, there is
no straightforward formulae in terms of cycle indices.
This is a consequence of the fact that we are no longer just counting
points on a double coset, but rather, we count those points on a double coset 
that satisfy some additional constraints.
To go further, a new approach is needed. The counting of ribbon graphs, refined 
by genus, has been considered in several papers in the mathematics 
literature, for example \cite{HarerZagier, jpv, nedela1, nedela2}.  
The method used in \cite{nedela1,nedela2} relies on the Burnside Lemma 
and a procedure of quotienting by automorphisms.  
In section 3, we develop 
this method in order to address the problem  of calculating a 
more refined counting function $\cN ( T_1, T_2, T_3)$ for embedded 
bi-partite graphs, 
which depends on three conjugacy classes $ T_1, T_2, T_3 $ of a
symmetric group $S_d$. We focus attention on connected graphs. 
 By employing the Burnside lemma, 
the counting is expressed by introducing a sum over an additional 
permutation which commutes with the permutation triple 
 describing the ribbon (or more general bi-partite) graph. 
This sum is simplified dramatically by employing Hall's theorem 
\cite{nedela,Liskovets1,Liskovets2}, which
constrains the commuting permutation to have all cycles of equal length $l$, 
thus generating a cyclic $Z_l$ subgroup of $S_d$.  
The  permutations $\gamma$ which commute with the triple 
$\{  \s_1, \s_2 ,\s_3 \}  \equiv L $ describing a graph form the automorphism group of the graph. The order of 
this automorphism group is the usual symmetry 
factor of large $N$ Feynman rules.  Given an automorphism $\gamma$  of a
 triple $L$ (which is also called a labelled graph), 
 we can quotient by  the automorphism to obtain a quotient graph 
$\bar L = \{ \bar \sigma_1 , \bar \sigma_2 , \bar \sigma_3 \}  $
 which has a number of edges  $ \bar d  = d / l $. 
The quotienting procedure also defines a branched covering map 
from the Riemann surface $ \Sigma $ 
supporting $L$ to the Riemann surface $ \bar \Sigma $ supporting 
$\bar L$. By covering space theory, there is an  associated 
 epimorphism from $ ( \bar \Sigma \setminus {\rm  punctures} )  $ to $Z_l$. 
There are branch indices associated with each cycle of the permutations 
$ \bar \s_i$. In this way the counting  is reduced to  the problem of
computing the number of quotient graphs and epimorphisms. 
There is also an  index distribution factor which arises when 
multiple cycles of the same length in the same $\bar \s_i$ have 
distinct branch indices. 

We use our refined counting formulae in section 4 to obtain explicit
results for a number of infinite families
of bi-partite graphs.  These results make contact with and extend results
already available in the mathematics
literature.  Some of the families we compute agree 
with previously available results 
on chord diagrams  \cite{chord} and one-vertex triangulations 
\cite{triangulation} thanks to non-trivial
identities involving sums of characters of permutation groups. We have checked 
the relevant identities with Mathematica, but we do not always have nice
derivations. Understanding the general structure of the identities 
is an interesting problem for the future. The agreement between ribbon graph
and chord diagram counting involves a general 
transformation that can be performed on ribbon graphs which results in 
an embedded graph having only trivalent vertices.

In section 5 we generalize our counting results
to Feynman graphs that have external legs. 
For quantum field theory applications, these external legs are
distinguishable, leading to the notion of
external-edge-labelled (EEL) graphs which we define. The EEL graphs can be
put into one-to-one correspondence with elements of a double coset. Again
the application of Hall's theorem
gives a dramatic simplification and ultimately leads to a formula for the
number of EEL graphs as a sum
over characters. We also introduce a second double coset, obtained by
treating the external legs as indistinguishable, which has applications 
in matrix model correlators. These two double cosets are related to each other 
via a quotienting by the permutation group of external edges.

Given the equalities between counting problems of Feynman graphs
with the counting of amplitudes in a string theory uncovered in \cite{MR2},
it is natural to explore the geometry
of the refined counting we have considered in this paper.
This question leads to a rather rich geometrical description which is
developed in section 6.
The geometrical description employs a three dimensional topological field
theory which counts maps
from membranes to a 3-manifold $\Sigma\times S^1$, where  $\Sigma$ is  a
3-punctured sphere. In section 7 we discuss our results and point out 
interesting directions in which they may be extended.

Since this paper links different areas of string theory, quantum
field theory and combinatorics, we provide a glossary in the Appendix A, which
should be useful to diverse readers. The remaining Appendices collect
detailed examples of the methods developed in the main body of the paper
and supply the technical details of a number of computations.

\section{Review : Permutation triples,  Bi-partite graphs  and  correlators } 

\subsection{ Matrix model correlators and bi-partite graphs }
\label{matbip}  

We review the description of Hermitian matrix model correlators
in terms of permutation triples explained in \cite{MR1}. For earlier relevant 
literature see \cite{bauitz,looijenga}. Observables can be parametrized by 
permutations, more precisely conjugacy classes, equivalently cycle structures of permutations. 
A multi-trace operator containing $d= 2n$ copies of matrix $\Phi$ determines a partition of $d$, 
or a conjugacy class in $S_d$. Call this conjugacy class $T_1$, let 
$\s_1$ be a permutation in that conjugacy class and define 
\bea\label{obsconj}  
\cO_{ T_1 } = \Phi^{i_1}_{i_{\s_1 (1) } } \cdots \Phi^{i_d}_{i_{\s_1 ( d ) } }
\eea
Wick contractions are pairings, corresponding to the conjugacy class 
$[2^n]\equiv T_2$, consisting of permutations with $n$ cycles of length $2$. 
  Using diagrammatic tensor space techniques,
  the following equation was derived  
\bea\label{corform}  
\langle \cO_{T_1} \rangle = { 1 \over |T_1| } 
\sum_{ \s_1 \in T_1 } \sum_{ \s_2 \in T_2 } \sum_{ \sigma_3 \in S_d  } 
\delta ( \sigma_1 \sigma_2 \sigma_3 ) N^{ C_{ \sigma_3 } }
\eea
in \cite{MR1}.
The $\delta ( \sigma ) $ is defined as being $1$ when its argument 
is the identity element of the permutation group, and $0$ for any other group 
element.
 We denote by $|T|$ the number of permutations in 
the conjugacy class $T$. If there are $t_i$ cycles of length $i$ 
in this conjugacy class, then this number is 
$$ { d! \over \prod_{i=1}^d i^{t_i} t_i! }.  $$ $C_{\s_3}$ 
is the number of cycles in the permutation $ \s_3$.  
The usual method of doing large $N$ calculations is to draw 
double line diagrams, or equivalently ribbon graphs, which are closely related to the above formula.  As an example consider the correlator 
$ \langle tr (\Phi^4 )  \rangle $ in the Gaussian Hermitian matrix model. 
There are three Wick contractions, which can be drawn using 
the double line notation, with lines indicating how the 
indices are identified when we use 
\bea 
 \langle \Phi^i_j \Phi^k_l \rangle = \delta^i_l \delta^k_j 
\eea 
for each contraction. 

%
%
%%%%%%%%%%%%%%%%%%
\begin{figure}[h]
\scalebox{1.0}{
\centerline{\includegraphics[height=5.5cm]{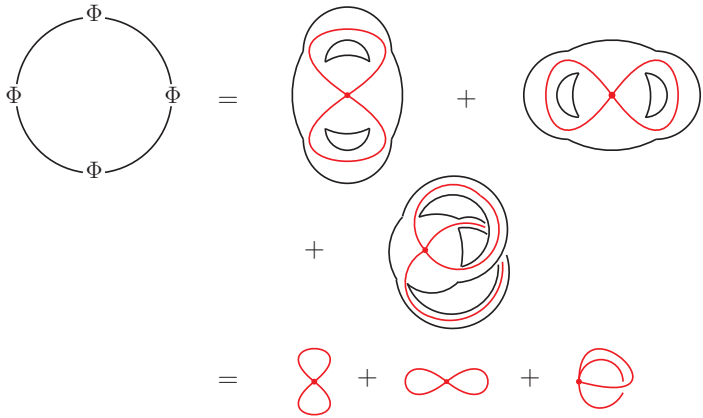} }
%\centerline{\includegraphics{triangle} } 
}
\caption{\it
Double-line diagrams and ribbon graphs  }
 \label{fig:dbrib}
 \end{figure}
%%%%%%%%%%%%%%%%%%%
%
%

\bea\label{WickCont}
\langle {\rm Tr}( \Phi \Phi \Phi \Phi ) \rangle =
\langle { \rm Tr } \contraction{}{ \Phi}{\Phi }{\Phi }
\contraction[2ex]{\Phi }{\Phi }{\Phi }{\Phi }
\Phi \Phi \Phi \Phi \rangle   
+  
\langle { \rm Tr } \contraction{}{\Phi }{}{\Phi}
\contraction[1ex]{\Phi\Phi }{\Phi}{}{\Phi}
\Phi \Phi \Phi \Phi\rangle   
+  
\langle { \rm Tr } \contraction[2ex]{}{\Phi}{\Phi \Phi }{\Phi}
\contraction[1ex]{\Phi }{\Phi}{}{\Phi}
\Phi \Phi \Phi \Phi  \rangle   
\eea

 Figure \ref{fig:dbrib} shows double-line diagrams corresponding to 
the Wick contractions  in (\ref{WickCont}). 
As shown in the Figure,   we can replace the double lines 
with a graph embedded on the two-dimensional surface. The first 
two graphs can be drawn without intersection
 on the sphere, while the third can be drawn 
without intersection on a torus. In fact the first two are isomorphic as ribbon 
graphs. The orientation of the underlying surface endows the 
vertices with a cyclic order. We can think of these graphs in 
close analogy to ordinary graphs, namely a collection of vertices 
and edges starting and ending on the set of vertices, but with the 
understanding that the vertices have a local cyclic symmetry rather than 
a complete permutation symmetry. Indeed this is how ribbon graphs are often 
defined ( see e.g \cite{looijenga}, \cite{LanZvo}).  This  description can be
 used to develop group theoretic algorithms for counting
ribbon graphs and their symmetry factors. 
Essentially the same group theoretic algorithms work for 
ordinary Feynman graphs as for ribbon graphs, with permutation groups 
and cyclic groups exchanging roles as the local symmetries of the 
vertices \cite{MR2}. These embedded graphs or ribbon graphs 
also  called ``maps" in some of the maths literature, e.g \cite{nedela}

The relation between the description of matrix model 
correlators in terms of permutation triples given in (\ref{corform})
and the one in terms of ribbon graphs can be understood very simply. 
This is done by sub-dividing the edges of the ribbon graph 
by the introduction of a new type of vertex. We can call 
all the previously existing vertices of the ribbon 
graph, black vertices, and the newly added vertices white vertices. 
The edges of the new graph (sometimes called half-edges
since they came from dividing edges into two) can be labelled 
$1 , \cdots , d $. A permutation $ \sigma_1 $ is defined 
by going round the black vertices and a permutation $ \sigma_2$ is 
defined by going round the white vertices, both done according to 
a fixed orientation on the surface.

\begin{figure}[h]
\scalebox{1.0}{
\centerline{\includegraphics[height=4.5cm]{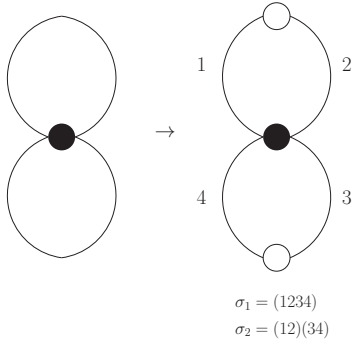} }
%\centerline{\includegraphics{triangle} } 
}
\caption{\it Ribbon graph to permutation
  }
 \label{fig:cs1s2}
 \end{figure}

 This is illustrated in Figure 
\ref{fig:cs1s2}. 

%
%
%%%%%%%%%%%%%%%%%%
%%%%%%%%%%%%%%%%%%%
%
%

We have written permutations in cycle notation, whereby 
$(1,2,3,4)$ denotes the re-arrangement 
 $ \{ 1 , 2 , 3 , 4 \} \rightarrow \{ 2 , 3, 4, 1 \} $.

\begin{figure}[h]
\scalebox{1.0}{
\centerline{\includegraphics[height=4.5cm]{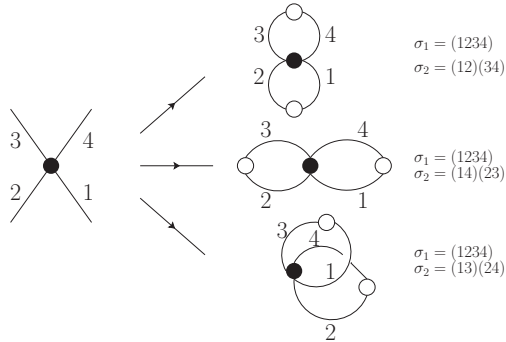} }
%\centerline{\includegraphics{triangle} } 
}
\caption{\it Wick contractions and permutations $\sigma_2$
  }
 \label{fig:ws1s2}
 \end{figure}

This description provides a useful way to think about the Wick contractions.
We can choose a labelling around the vertices, which 
determines $\sigma_1$. Then the different Wick contractions 
correspond to different choices of $\sigma_2$. This is illustrated 
in Figure \ref{fig:ws1s2}.  
%
%%%%%%%%%%%%%%%%%%
%%%%%%%%%%%%%%%%%%%
%
%

The fact that the first two Wick contractions give the same ribbon 
graph (and the same power of $N$) follows from the fact that 
there is a permutation $ \gamma  = ( 1,2,3,4)$ which conjugates 
the first pair $ \{ \sigma_1 , \sigma_2 \} = \{ ( 1,2,3,4 ) , ( 1,4)( 2,3) \} $
to the second pair $ \{ \sigma_1' , \sigma_2' \} =
 \{ ( 1,2,3,4 ) , ( 1,2) ( 3,4) \} $.
Given the two permutations, we can multiply them to get $ \sigma_3 = ( \sigma_1 \sigma_2)^{-1} $ so that 
\bea 
\sigma_1 \sigma_2 \sigma_3 = 1 
\eea
This third permutation contains information about the faces
of the embedded graph. The number of cycles in $\sigma_3$, 
denoted $C_{ \sigma_3} $ is equal to the number of faces. 
This, as expected from the double line notation, is the power of $N$
we get from double line diagrams and this is indeed consistent with 
(\ref{corform}) for the correlator. 
Permutation triples  related by  conjugation 
\bea\label{equivtriples}  
\sigma_i' = \gamma \sigma_i \gamma^{-1} 
\eea 
correspond to the same ribbon graph, as in the example above.  
Ribbon graphs can be identified 
with equivalence classes of these permutation triples. 
A triple is also called a labelled graph $L = \{ \s_1 , \s_2 , \s_3 \}$
and the ribbon graph or equivalence class is denoted $[L]$.

The genus of the Riemann surface supporting the ribbon graph 
can be inferred directly from the permutations. It is given by 
the Riemann-Hurwitz relation
\bea\label{RH}  
( 2g - 2 ) = d - C_{ \s_1 }  - C_{  \s_2  }  - C_{  \s_3 }  
\eea
where $C_{\s_i}$ is the number of cycles in the permutation 
$\s_i$. This is  understood using the fact
that  the data of permutation triples defines a holomorphic map 
from the Riemann surface to a sphere $\mP^1$  with 3 branch points. 
This is called a Belyi map. Such maps have a special role in number theory, 
since they can be defined over algebraic number fields. Ribbon graphs drawn 
on the Riemann surface are also called Dessins d'Enfants. 
This connection to Belyi maps was used in \cite{MR1} to argue 
that the Hermitian matrix model is dual to a string theory 
of holomorphic maps from worldsheet to $\mP^1$. This has been 
discussed at the level of the standard topological A-model string
in \cite{gopak}.  

An important property of  a ribbon graph is whether it is connected or not. 
This is read off from the permutation triple by asking whether 
they generate a subgroup of $S_d$ which acts transitively on the 
set $\{1, 2 , \cdots , d \}$. For any pair of elements $i,j$ in this set, 
a transitive subgroup contains a permutation which acts on $i$ to give $j$. 

We can generalize the discussion to  let the white vertex structures be of any 
 conjugacy class. This is useful in the complex matrix model. 
In this case, holomorphic and anti-holomorphic observables 
can be parametrized by conjugacy classes of permutations $T_1,T_2$ as 
in (\ref{obsconj}) and the answer expressed in terms a sum of permutations 
tracking the Wick contractions \cite{cjr}. The expression can be simplified,
eliminating the Wick contraction permutations \cite{tom},  to give 
\bea\label{complcorform}  
\langle \cO_{ T_1 } \cO_{ T_2}^{\dagger} \rangle 
= { d! \over |T_1| |T_2| } \sum_{ \sigma_{1} \in T_1 } \sum_{ \sigma_2 \in T_2 } \sum_{ \sigma_3 \in S_d }  \delta ( \sigma_1 \sigma_2 \sigma_3 ) N^{ C(\sigma_3) } 
\eea
Here $d$ is the number of $\Phi$ in the holomorphic observable  $\cO_{ T_1 }$
or the number of $\Phi^{\dagger}$ in the anti-holomorphic 
observable  $\cO_{ T_2}^{\dagger}$. 

In  (\ref{corform}) or (\ref{complcorform}) we note that, 
while the sums over $\s_1 , \s_2 $ run over fixed 
conjugacy classes $T_1, T_2 $,  the sum over $\sigma_3$ 
runs over the whole permutation group. We can define a refined 
counting  where the conjugacy class $T_3$ of $\sigma_3$ is fixed. 
This corresponds to fixing the structure of the faces of the ribbon graph. 
If $T_3$ has $p_1$ cycles of length $1$, $p_2$ cycles of length $2$ 
etc. then the ribbon graph has $p_1$ faces bordered by two edges and 
one black,  one white vertex, $p_2$ faces bordered by four edges and two   
black, two white vertices etc.

Indeed, we may define  $\cZ ( T_1 , T_2 , T_3 ) $ 
\bea\label{deltatriple}  
\cZ ( T_1 , T_2 , T_3 )  = \sum_{ \s_i \in T_i } \delta ( \s_1 \s_2 \s_3 ) 
\eea 
 Conjugacy classes form an algebra and $\cZ( T_1, T_2, T_3)$ 
are the structure constants of the algebra. They can be computed using 
Symmetrica Online \cite{symmetrica-classsum}.  
It also has a very useful expansion in terms of characters 
$\chi_R ( \s_i)$ for group element $\s_i \in T_i$ in representation 
$R$ associated with Young diagrams, which allows us to write  
\bea\label{charsumcz}  
\cZ ( T_1 , T_2, T_3 ) =  { |T_1| |T_2|  |T_3| \over d! } 
\sum_{ R \vdash d } { \chi_R ( \s_1 ) \chi_R ( \s_2 ) \chi_R ( \s_3 ) \over d_R } 
\eea
We will use  the Murnaghan-Nakayama (MN) Lemma (see \cite{fultonharris}) 
to  compute this sum for certain infinite classes of conjugacy classes
in Appendix  \ref{charactersApen}. 

The quantity $\cZ( T_1 , T_2 , T_3) $ is an observable in large $N$ 
two dimensional Yang Mills theory \cite{grta,cmr} and has an interpretation 
in terms of the dual string theory. It is related to the insertion of 
the Wilson loop observables  at the 3 boundaries of a 3-holed sphere. 
 The subset of permutations 
$\gamma \in S_d $ which leave all three $\s_i$ fixed under conjugation,
form a subgroup, called the automorphism group of the triple. In the context 
of the covering space interpretation, this is a subgroup of 
homeomorphisms $\phi $  of the covering space which leave the covering map $f$
unchanged, i.e $ f = f \circ \phi$. The size of this Automorphism group 
is also the symmetry factor which appears in the denominator in large $N$  Feynman 
rules. The quantity $\cZ $ is 
 a sum over ribbon graphs, weighted with inverse symmetry factor.
\bea 
\cZ ( T_1 , T_2, T_3 ) = d! \sum_{ \hbox{ ribbon graphs}  } { 1 \over 
| \hbox{ Aut ( ribbon graph ) } |   } 
\eea

 There is another closely related quantity where we count 
each ribbon graph, or equivalence class of permutation triples, 
with weight one. Define 
\bea 
\cN ( T_1 , T_2 , T_3 )  & =& \hbox{  Number of equivalence classes of triples }  \cr  
                                      & =& \hbox{ Number of ribbon graphs with specified vertex and face structure     } \cr 
&& 
\eea
This is clearly the fundamental combinatoric element in perturbative large $N$ 
QFT computations, since we need to be able to enumerate the graphs and then 
compute corresponding Feynman integrals. This refined 
counting problem, depending on the choice of 3 conjugacy classes $T_1, T_2 , T_3 $ will be the main object of 
interest in this paper. We will extend existing
  mathematical techniques from graph theory for application to 
this physics problem. These will be helpful not only in counting 
but also construction of  these graphs. A related and somewhat simpler counting 
problem is to fix $T_1 , T_2$ and sum over all $T_3$, giving 
counting functions $ \cN ( T_1 , T_2)$ which depend on 
only two conjugacy classes specifying the valencies of the 
black and white vertices.  All the elements in a conjugacy class 
$T_i$ can be obtained by fixing a permutation $\hat \sigma_i \in T_i $
and writing 
\bea 
\sigma_i = \alpha_i \hat \sigma_i \alpha_i^{-1} 
\eea 
The $\alpha_i $ are permutations in $S_d$. They produce the same 
$\sigma_i$ when multiplied on the right by  elements in 
the subgroup $H_i$ which commute with $ \hat \sigma_i $. 
So the set of $\sigma_i$ can be identified with the coset 
$ G / H_i$. The counting of $ \cN( T_1 , T_2 ) $ can 
be expressed in terms of the double coset $ H_1\setminus G / H_2$
and this leads to formulae in terms of cycle indices of $H_1 , H_2$ 
(see \cite{MR2} and references therein).  We can also fix the black vertex structure, sum over white vertex structure, while fixing the face structure, and the same type of techniques will work. For example one 
can compute the number of triangulations  : where 
the cycles of $T_3$ are all of length $3$ \cite{vidal}. 

The quantities $ \cZ , \cN$ are symmetric under 
permutations of the three arguments : 
\bea 
&& \cZ ( T_1 , T_2, T_3 ) = \cZ ( T_3 , T_2, T_1 ) = \cZ ( T_2 , T_1, T_3 ) \cr 
&& \cN ( T_1 , T_2, T_3 ) = \cN ( T_1 , T_2, T_3 ) = \cN ( T_2 , T_1, T_3 )
\eea 
These can give rise to non-trivial relations between different types 
of matrix model correlators or between apparently different  
graph counting problems. For example the exchange of $T_1, T_3$ is 
face-vertex duality.

 The quantity $ \cZ ( T_1, T_2 , T_3 ) $ counts what are called 
 {\it labelled maps} in the mathematics literature 
 (i.e labelled embedded graphs in the terminology of this paper) 
  and   $\cN ( T_1 , T_2 , T_3)$ counts { \it unlabelled maps} or {\it unrooted maps}
  (i.e embedded graphs in our terminology). For the connected 
case, the number of labelled maps is $(d-1)! $ times the number 
of what are called {\it rooted maps}. The literature on  rooted/labelled maps 
is more extensive than that on unrooted/unlabelled maps. 

For computations in scalar field theory the quantities  
$ \cN ( T_1 , T_2 , T_3) $ directly give the desired Feynman graphs 
of large $N$. For more general field theories, a classification of 
these graphs provides a starting point, upon which additional data associated with spin and flavour quantum numbers can be incorporated. Indeed in standard computational approaches to Feynman rules  (without large $N$)  the graphs of scalar field theory are the starting point \cite{Nogueira}. Detailed 
studies of $ \cN ( T_1, T_2, T_3 )$ are therefore of very general interest 
for large $N$ computations. In the next sub-section we will show that 
amplitude computations, even without large $N$, involve this counting problem : 
here the source of the local cyclicity will not be double-line diagrams but 
the nature of MHV vertices. 

\subsection{ MHV rules and ribbon graphs}\label{sec:PermsAndMHVrules}

Scattering amplitudes in Yang-Mills theories can be decomposed according to their color structures, 
\bea \label{amplitude}
\mathcal{A}_n(\{ p_k, a_k \} ) 
&=& \sum_{\sigma \in S_n/Z_n} N {\rm Tr}(T^{a_{ \sigma(1)} }, \ldots, T^{a_{\sigma(n)}}) 
A(p_{ \sigma(1)} ,\dots, p_{\sigma(n)} ) 
\\ \nonumber \cr
&+& \sum_{\sigma \in S_n/S_{n;c} } 
{\rm Tr}(T^{a_{ \sigma(1)} }, \ldots,T^{a_{ \sigma(i)} }) 
{\rm Tr}(T^{a_{ \sigma(i+1)} }, \ldots,T^{a_{ \sigma(n)} }) \\ \nonumber \cr
&\times & A(p_{ \sigma(1)} ,\ldots, p_{ \sigma(i)}  ; p_{ \sigma(i+1)} , \ldots, p_{ \sigma(n) } )   \cr 
&+&   \cdots ,
\eea
where $T^{a}$ is a $U(N)$ generator, and $p_i$ is momentum. Furthermore $Z_n$ and $S_{n;c}$ are the symmetries of the single trace and double trace. They also leave the kinematics parts, $A(p_{ \sigma(1)} ,\dots, p_{ \sigma(n)} )$ and $A(p_{ \sigma(1)} ,\ldots, p_{ \sigma(i)}  ; p_{ \sigma(i+1)} , \ldots, p_{ \sigma(n) } )$ invariant.
The ellipses indicate possible higher multiple trace terms. 
At tree-level the amplitudes have only single trace structure, whereas one-loop amplitudes can also have double-trace structure and more complicated trace structures appeared at higher-loop level. Each color structure is associated with a color ordered partial amplitude, for instance 
$ A(p_{ \sigma(1)} ,\dots, p_{ \sigma(n)} )$ and $A(p_{ \sigma(1)} ,\ldots, p_{ \sigma(i)}  ; p_{ \sigma(i+1)} , \ldots, p_{ \sigma(n)} )$ in the above equation. It is convenient to classify the amplitudes in terms of the number of negative helicity gluons. For a supersymmetric theory, that is what we will focus on, the first non-trivial amplitude is the one with two negative helicity gluons, which is often called maximally-helicity-violating (MHV) amplitude in the literature. Similarly the amplitudes with $k+2$ negative helicity gluons are referred to as N$^k$MHV amplitudes. Any $n$-point tree-level MHV amplitude was first obtained by Parke-Taylor in a very simple and beautiful form~\cite{Parke:1986gb}
\bea
\mathcal{A}^{\rm MHV}_{n}(\{ p_k, a_k \} )
&=& \sum_{\sigma \in S_n/Z_n} {\rm Tr}(T^{a_{ \sigma(1)} }, \dots, T^{a_{ \sigma(n)} }) 
  {\delta^8(\sum_i Q_i) \delta^4(\sum_i p_i) \over \langle 1 2\rangle \dots \langle n\!-\!1 n\rangle \langle n 1\rangle} \, ,
\eea 
where $p_i$ is momentum and ${Q^{A}_i}_{\alpha} = {\lambda_i}_{\alpha} \eta^A_i$ is often referred to as super momentum. Here $\eta^A$, with $A=\{1, 2, 3, 4 \}$, is the fermionic variables for $\mathcal{N} = 4$ supersymmetry. We have applied the  helicity spinor formalism and the on-shell $\mathcal{N}=4$ notation to write the super amplitudes in a compact form. For a recent review on the notations and the subject of  scattering amplitudes see e.g.~\cite{Feng:2011gc}. Let us only mention that the holomorphic spinor $\lambda$ and anti-holomorphic $\tilde{\lambda}$ are used to solve the massless on-shell condition $p^2 =0$, which are defined as
\bea
p_i = \lambda_i \tilde{\lambda}_i \, .
\eea

 The MHV rules give a powerful tool for calculating non-MHV scattering amplitudes. Inspired by the twistor string formalism of $\mathcal{N}=4$ SYM \cite{Witten:2003nn}, the MHV rules were originally formulated at tree-level \cite{CSW}, and later were generalized to loop-level \cite{BST}. The rules state that any scattering amplitude in $\mathcal{N}=4$ SYM can be computed by gluing super MHV vertices together with scalar propagators. Any amplitude  can be written as 
\bea
&& \mathcal{A}_n(\{ p_k , a_k \} )  = \int \prod_i { d^4 \eta_i d^4 P_i \over P_i^2 } 
 \mathcal{A}^{\rm MHV}_{n_1}(\{ p_{k_1} , a_{k_1}  \} )  
  \mathcal{A}^{\rm MHV}_{n_2}(\{ p_{k_2} , a_{k_2}  \} ) 
   \dots  \mathcal{A}^{\rm MHV}_{n_m}(\{ p_{k_m} , a_{k_m}  \} ). \cr
&& 
\eea
where the $P_i$ are the internal  momenta; the $p_{k_i } $  include 
internal and may have external momenta.  Since the propagators are off-shell, the spinors for the internal propagators appearing  in the Parke-Taylor formula have to be defined by  an off-shell prescription. The definition of the off-shell continuation was realized in~\cite{CSW} by introducing an arbitrary reference spinor $\tilde{\xi}$, 
\bea
\lambda_{i  \, \alpha } = P_{i  \, \alpha, \dot{\alpha}} \tilde{\xi}^{\dot{\alpha}}  \, ,
\eea 
and the final result, after summing over all the MHV diagrams, is independent of $\tilde{\xi}$, reflecting gauge invariance. For a understanding of the MHV rules at the level of a Lagrangian see~\cite{Ettle:2006bw}. Let us finally mention briefly the color structure part of MHV rules. 
The $U(N)$ generator $T^a$ satisfies
\bea \label{unrelation}
{\rm Tr}(T^a T^b) = \delta^{ab}\, , \qquad 
\sum_a (T^{a \, \bar{i}_1}_{\,\, \, j_1} T^{a \, \bar{i}_2}_{\,\,\, j_2}) 
= \delta^{\bar{i}_1}_{j_2} \delta^{\bar{i}_2}_{j_1} \, .
\eea 
The color structure of MHV diagrams can be obtained from those relations by gluing MHV vertices together~\cite{LW}. 
MHV rules were used to prove the standard relation between single trace partial amplitudes and double trace amplitudes in equation (\ref{amplitude}) at one-loop level~\cite{LW}. For further details on MHV rules and its application  see, e.g. the review ~\cite{Brandhuber:2006vh}. For the purposes of our discussion, what we want to stress is the fact that any $n$-point MHV vertex has a $Z_n$ cyclic symmetry.
Thus it is natural to formulate MHV diagram counting problems in terms of our framework by associating each MHV vertex with a cycle in a  permutation $\sigma_1$, as we will illustrate in the following example. 

%
%
%
%%%%%%%%%%%%%%%%%%
\begin{figure}[h]
\scalebox{1.0}{
\centerline{\includegraphics[height=4.5cm]{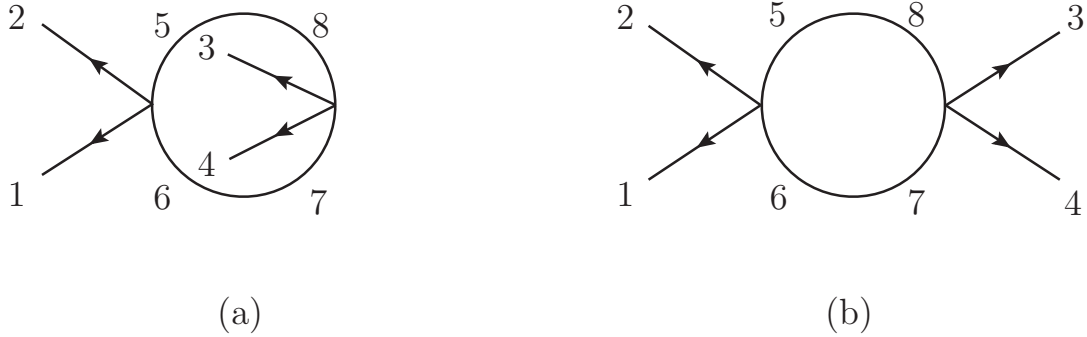} }
%\centerline{\includegraphics{triangle} } 
}
\caption{\it
MHV diagrams with two MHV vertices and four-point external legs, where Figure {\rm (a)} has a double trace structure and Figure {\rm (b)} has a single trace structure. }
 \label{fig:mhv}
 \end{figure}
%%%%%%%%%%%%%%%%%%%
%
% 
Here we consider one-loop MHV diagrams with four external legs. Label the external legs $1,2,3,4$, as in Figure \ref{fig:mhv}. 
We have two types of diagram, which are described by the permutation pairs,
\bea 
\sigma_{1}^{(a)}  & = &  ( 1256) (78 34 ) \cr 
\sigma_{2}^{(a)}  & = &   (57) ( 68)  
\eea 
and 
\bea 
\sigma_{1}^{(b)}  & = &  ( 1256) (78 34 ) \cr 
\sigma_{2}^{(b) }   & = &   (58) ( 67)  \, .
\eea
The trace structures of the MHV diagrams are determined by the products of the permutations, $ \sigma_1 \sigma_2 $,
\bea
&& {\rm (a)} ~~ : ~~  ( \sigma_{3}^{(a)})^{-1}   = \sigma_{1}^{(a) }  \sigma_{2}^{(a)}  = (1276) (3458) \cr 
&& {\rm (b)} ~~ : ~~ ( \sigma_{3}^{(b)} )^{-1}  = \sigma_{1}^{(b)}  \sigma_{2}^{(b)}  = (128346) (57)  
\eea
If we drop the indices which appear in the pairings $\sigma_2$'s, 
and denote this operation as $ D ( . ) $, then we have 
\bea 
&& {\rm (a)} ~~ : ~~  D (  \sigma_{3}^{(a)} ) = (12)(34) \cr 
&& {\rm (b)} ~~ : ~~    D (  \sigma_{3}^{(b)}  ) = ( 1432) () \, ,
\eea 
where we interpret the empty cycle as $N$.  The result shows that the trace structure of case (a) is a double trace, ${\rm Tr}(T^1 T^2){\rm Tr}(T^3 T^4)$, 
while the trace structure of the case (b) is a single trace, ${\rm Tr}(T^1 T^2 T^3 T^4)$ with a factor of $N$. The result is consistent with what one would obtain by applying the $U(N)$ generator relations, i.e. equation (\ref{unrelation}).  The cyclic nature of the super MHV vertices thus allows a treatment of the combinatorics of  MHV diagrams in terms of ribbon graphs and hence  permutation triples, which neatly encode the essential 
physics of trace structures. 

In fact in any QFT problem involving vertices with a cyclic symmetry, another example being 
non-commutative field theories (see e.g. \cite{minseib}) ,  ribbon graphs will be relevant.

\subsection{ Bi-partite graphs and Feynman graphs of QED } 

\begin{figure}[h]
\scalebox{1.0}{
\centerline{\includegraphics[height=4.5cm]{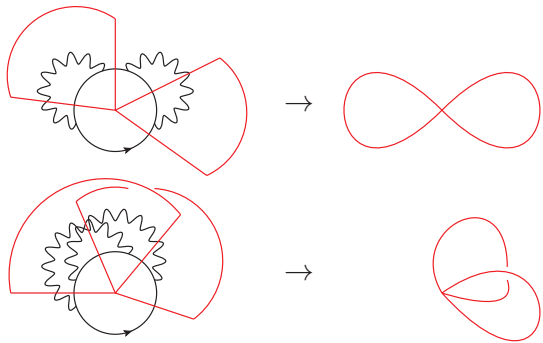} }
%\centerline{\includegraphics{triangle} } 
}
\caption{\it QED Reynman graphs to ribbon graphs }
 \label{fig:QED2}
 \end{figure}

A somewhat surprising appearance of ribbon graphs
is in the context of vacuum graphs of QED, i.e $U(1)$ gauge theory 
coupled to a fermion or a complex scalar. The key reason
for this is illustrated in Figure \ref{fig:QED2}  
which shows how to construct ribbon graphs corresponding
to two vacuum diagrams. Inside each fermion loop we draw a vertex. 
We draw an edge going out to every arc lying between 
photon vertices. Since the matter loops are equipped with an orientation, 
the edges coming out of the newly introduced vertex inherit 
a cyclic structure necessary for ribbon graphs. This also allows us
to determine how the different edges are tied together. 
The correspondence between vacuum graphs of QED and ribbon graphs 
was found in \cite{MR2} by setting up the counting of the 
QED graphs in terms of permutations, and simplifying the resulting 
formulae to find a ribbon graph counting problem. This correspondence 
with ribbon graphs is distinct from the usual large $N$ gauge 
theory story, the local cyclic symmetry of the vertices arising
from the directed matter lines as opposed to the colour-lines 
describing matrix indices. 

%
%%%%%%%%%%%%%%%%%%
%%%%%%%%%%%%%%%%%%%
%
%

\subsection{ Bi-partite graphs on torus and AdS/CFT  }\label{sec:bipAds}  

Bi-partite graphs on a torus also appear in the description of 
gauge theories living on 3-branes transverse to toric 
Calabi-Yau singularities \cite{FHKVW}.  The edges of the 
graph correspond to chiral superfields in the theory. 
Black vertices correspond to positive terms in the superpotential. 
White vertices correspond to negative terms. 
Again we can describe the bi-partite graph in terms 
of a permutation triple, which also yields a convenient 
way to read off zig-zag paths \cite{jrr} which play a central role 
in relating the geometry to the gauge theory (see \cite{yam} for a review
of the subject).   First label the edges $\{ 1 \cdots d \} $. Read off 
permutations $ \s_1 , \s_2 $ by going in a fixed direction, say 
anti-clockwise around the black and white vertices, and obtain 
$ \s_3 = (\s_1 \s_2)^{-1} $ by group multiplication in $S_d$. 
Again the cycle structures of these three permutations 
are related to the genus of the surface (in this case genus one)
by the Riemann-Hurwitz formula. It is important for this 
that the two permutations are read off by going round the black and white 
vertices according to the same orientation, even though 
the construction of the superpotential uses a reading of the 
edges around black and white vertices in opposite directions.

Yet other appearances of ribbon graphs occur 
 in classifying subgroups of $SL(2 , \mZ )$ \cite{yang}. 

These multiple appearances of ribbon graphs in physics make 
it extremely  important  to understand their counting in generality
as well as the geometrical aspects  of this counting.

\section{ Quotienting bi-partite graphs and 
going back  via epimorphisms }\label{QuotsAndRev}

Bi-partite graphs can be counted in terms of 
equivalence classes of triples $(\s_1 , \s_2 , \s_3 ) $ in 
$S_d$ where $ ( \s_1' , \s_2' , \s_3' ) $ is equivalent to 
 $(\s_1 , \s_2 , \s_3 ) $ when 
\bea
\s_i' = \gamma \s_i \gamma^{-1}  
\eea
for some $\gamma \in S_d$. We are interested in counting 
with $ \s_i \in T_i$, with conjugacy classes $T_i$ fixed. 
Each triple $ \s_1 , \s_2 , \s_3 $ describes what is called 
a { \it labelled graph}. So we will write $ L = \{ \s_1 , \s_2 , \s_3 \} $.
These equivalence classes under $S_d$ conjugation are 
{\it unlabelled graphs}. We will use $ T ( L ) \equiv \{ T_1 , T_2 , T_3 \} $
to denote the conjugacy classes of the three permutations. 
And we use $[L] $ to denote the $S_d$ equivalence class containing $L$.
It is clear that $ T ( L ) = T ( [L] )$.

The quantity $\cZ( T_1 , T_2 , T_3) $ in (\ref{deltatriple})
is counting labelled graphs. The permutations in $S_d$ which leave 
a given labelled graph $L$ invariant form a subgroup $ \Aut ( L ) $. 
If $L, L'$ belong to the same equivalence class, then 
$\Aut (L)$  is conjugate to $\Aut ( L')$ by an element of $S_d$. 
This means that the order of the group, $|\Aut ( L ) |$ is only a function 
of $[L]$. We may write 
\bea 
| \Aut ( L )| = | \Aut ( [L] )| 
\eea
The number of labelled graphs in an $S_d$ equivalence class is equal to the 
\bea 
{ d! \over | \Aut ( [L] ) }  
\eea
So the partition function $Z( T_1 , T_2 , T_3) $ 
is a sum over equivalence classes $[L]$ weighted by the inverse order of 
the automorphism group. 
\bea 
Z( T_1 , T_2 , T_3)  = \sum_{ [L ] : T ( [L] ) = \{ T_1 , T_2 , T_3 \}  } 
{ 1 \over | \Aut ( [L] ) | } 
\eea

By the Burnside Lemma, these equivalence classes are counted by 
\bea 
&&  \cN ( T_1 , T_2 , T_3 ) 
=    \sum_{ [L ] : T ( [L] ) = \{ T_1 , T_2 , T_3 \}  } 
1                    \cr 
&& = { 1 \over d! } \sum_{ \gamma \in S_d  } \sum_{ \substack{
 \s_i \in S_d :\\ \s_i \in T_i  }  } 
 \delta ( \s_1 \s_2 \s_3 ) \delta ( \gamma \s_1 \gamma^{-1} \s_1^{-1} ) 
 \delta (  \gamma \s_2 \gamma^{-1} \s_2^{-1} )    
\eea
Denoting labelled graphs as $L$ we can write 
\bea 
 \cN ( T_1 , T_2 , T_3 ) = { 1 \over d! } \sum_{ \gamma \in S_d} \sum_{ L } 
 < \gamma  , L > 
\eea
 where $ < \gamma , L > $ is $1$ if $\gamma $ is an automorphism of  $L$
and zero otherwise.

Now for connected embedded graphs, Hall's theorem states that
$\gamma $ is a permutation which has cycles all of equal length. 
For completeness, we explain the argument in Appendix \ref{sec:Hall}. 
This means that the sum can be written 
as 
\bea 
\cN ( T_1 , T_2 , T_3 )  =  { 1 \over d! }
\sum_{ l | d  } \sum_{ \gamma \in [l^{\bar d}] } < \gamma , L > 
\eea
We are summing over divisors $l$ of $d$, with $\bar d  l = d$. 
For each $l$, we sum over permutations $\gamma $ which have $\bar d $ 
cycles of length $l$. It is easy to show that the number of labelled 
graphs fixed by $\gamma$ only depends on the conjugacy class of $\gamma$. 
If $\gamma $ fixes $ \sigma_i$ under conjugation action, then 
$\mu \gamma \mu^{-1} $ fixes  $ \mu \sigma_i \mu^{-1} $.  It follows : 
\bea
\cN ( T_1 , T_2 , T_3 )  && =  { 1 \over d! }
\sum_{ l | d  }     { d! \over l^{\bar d}  { \bar d} ! }  \sum_{ L }   <
\gamma , L > \cr
&& = \sum_{ l | d  }   { 1 \over l^{\bar d}  { \bar d} ! } \sum_{ L }  <
\gamma , L >
\label{useHall}
\eea
Here $\gamma $ is a fixed permutation in the conjugacy class 
$[l^{ \bar d } ] $, and we included a factor which is the
 size of the conjugacy class. For any $L= \{ \s_1 , \s_2 , \s_3 \} $
 which has $<\gamma , L > =1$,
we can generate a list by conjugating $L$ with permutations
in $S_d$ which leave $\gamma$ invariant under conjugation.
These permutations form a subgroup $S_{ \bar d }  [Z_l ] = S_{ \bar d }  \ltimes Z_l^{ \bar d } $.  
So the set of $L$'s  which are fixed by $\gamma $ form orbits 
of $S_{ \bar d }  [Z_l ]$. 

Given any $( \gamma , L ) $ pair, we can form a quotient to obtain 
$\bar L = \{ \bar \s_1 , \bar \s_2 , \bar \s_3 \} $. 
Choose a fixed permutation $\gamma$ with cycle structure 
$[l^{\bar d } ] $. Denote by $S_1, \cdots , S_{\bar d } $  the 
sets of integers in the cycles of $\gamma$. 
Denote the integers inside 
each cycle as $S_{a, b }$. The second index can be viewed as 
an element of  $Z_l$. Given any $L$ we can get a 
triple  of permutations of $\{ S_1 , \cdots , S_d \}$
by a quotienting construction.  
This triple $\bar \sigma_{i}$ is defined by 
\bea 
\bar \sigma_{i} ( S_a )  = \hbox{ cycle containing }  \sigma_i ( S_{a, b } ) 
\eea
This is well-defined, independent of the choice of $b $ \cite{nedela}. 
The quotienting construction gives, for each cycle of 
$\bar \sigma_i$, an index. Write 
\bea 
\bar \sigma_i = \prod_j  (\bar \sigma_i)_j 
\eea
with $j$ running over the cycles of $\bar \sigma_i$. 
Pick an element $S_{a  } $ in the cycle  $(\bar \sigma_i)_j $. 
Pick an element $S_{a, b}$ in the set  $S_a$. Define 
\bea\label{indexofcycle}  
\nu_{i,j} = { \hbox { Length of cycle in $\s_i $ containing   $S_{a,  b}$ } 
 \over  \hbox{ Length of cycle $(\sigma_i)_j$ } }  
\eea
These numbers are independent of the choices and always 
integral \cite{nedela}. 
So $\nu ( \bar{\sigma}_i ) $ is a set of positive integers $\nu_{i,j}$  assigned to
the cycles of  $\sigma_i$. Define  $\nu ( \bar L  ) = \{ \nu ( \bar \s_1 ) , \nu ( \bar \s_2 ) ,   \nu ( \bar \s_3 )  \}  $.  
More details and examples of the quotienting construction are described 
 in Appendix \ref{Appsec:quotienting}.

There is an  action of $S_d$ on the set of all pairs 
$(\gamma , L ) $ by simultaneous conjugation. The subgroup 
which fixes a given $\gamma $ is $S_{ \bar d } [ Z_l ] $. 
So the set of all $ L$'s fixed by a given $\gamma $
forms orbits of this $S_{ \bar d } [ Z_l ] $. 
The $Z_l^{ \bar d } $ subgroup of this stabilizer group
acts on the $\bar d $ indices $\{ b_1 , \cdots , b_{\bar d  }  \} $. The 
$S_{ \bar d } $ subgroup acts on the $a$ index of $S_{a,b_a}$. 
We can consider orbits of the $Z_l^{ \bar d}$ on the set of 
$L$'s fixed by a given $\gamma$.

We can  see that the  diagonal $Z_l$ acts trivially on the $L$'s
but the quotient $Z_l^{ \bar d}/ Z_l $  acts non-trivially, giving orbits of 
size $l^{ {\bar d } -1 }  $.  Indeed, the diagonal $Z_l$ 
is generated by $\gamma $ itself, so that is an automorphism of  $L$ by assumption. 
Given something of the form $( g_1 , g_2 , \cdots , g_{ \bar d } )$ 
in $Z_l^{ \bar d } $ its action on the set of $L$'s will 
be the same as the action of $  (1 ,  g_2 g_1^{-1} , g_3 g_1^{-1} , \cdots ,
 g_{ \bar d }  g_1^{-1}  ) $, since 
$ ( g_1^{-1} , g_1^{-1} , \cdots , g_1^{-1} )  $ acts trivially. 
However   \break
$  (1 ,  g_2 g_1^{-1} , g_3 g_1^{-2} , \cdots ,
 g_{ \bar d }  g_1^{-1}  ) $ acts on the set $ \{ 1 , \cdots , d \} $
by fixing one subset of $l$ elements, while moving the remaining 
elements in cycles of length $l$. By Hall's theorem, this cannot be 
in the automorphism group  of $L$, so this means that the $Z_l^{\bar d} $ subgroup 
of the $S_{\bar d } [ Z_l ] $ stabilizer of 
a fixed $\gamma$  generates orbits of size $l^{\bar d -1 }$ 
in the set of all $L$'s stabilized by $\gamma$.

Since there is an $S_{\bar d } $ action 
on the $L$'s fixed by a given $\gamma$,  the list of 
 $ ( \bar L , \nu ( \bar L) )$ will contain all the labelled 
$ \bar L $ ( i.e complete orbits of  $S_{\bar d } $). 
Each labelled $\bar L $ will have a multiplicity which 
includes the factor $l^{\bar d -1 } $ 
 identified above, along with two additional factors : the {\it epimorphism factor} 
 and the {\it index distribution factor}.  The epimorphism factor arises because the 
 quotienting procedure for getting a labelled $\bar L $
  from the labelled $L$  gives rise to additional data associated with 
  each cycle of the permutations in   $\bar L$ : namely  a 
  group element in $Z_l$. There are also $Z_l$ group elements associated 
  with non-trivial cycles on the Riemann surface supporting $ \bar L$.  
 The group elements  associated with cycles are denoted $e_k$ 
 and they obey $e_k^{\nu_k } =1 $, where $\nu_k$ is the index associated with that cycle. 
Denoting by $g$ the genus of the quotient ribbon graph, there are $2g$ non-trivial 
cycles on its supporting Riemann surface, with corresponding group elements $a_i , b_i$. 
The Riemann surface supporting $L$ is a branched cover of the one 
supporting $ \bar L$, so by covering space theory, we have  
\bea \label{epieqn} 
\prod_{i=1}^{g}  [a_i , b_i]  \cdot \prod_k e_k =1  ~~~ ; ~~~ e_k^{\nu_k} =1 
\eea
which will be called epimorhism equation. There is thus an  epimorphism from fundamental group of Riemann surface minus points to $Z_l$. 
In the appendix we explain, with examples, how the epimorphism is read off from $L$ and $ \bar L$. 

There is a simple class of cases, where  like cycles in $\bar L$ do  not have distinct multiplicities. 
Then the counting formula takes the form  
\bea\label{simpform}  
&& \cN ( T_1 , T_2 , T_3 )  = \sum_{ l | d  }   { 1 \over l  { \bar d} ! }   
 \sum_{ [ \bar L ] } 
\hbox{ Number of labelled $\bar L $ in equiv class  $[ \bar L ]$ } \cr 
&&\qquad \qquad \qquad  \times \hbox{ Number of epimorphisms }
\eea

This formula is not adequate in the most general case of refined counting. 
The easiest way to understand this is to look at
the first example in \ref{App:exampsNCF}. A subtlety arises when we 
have distinct indices for cycles of the same length in a given $\bar \s_i$.  
We have previously defined a set of integers $\nu_{i,j}$ 
associated to each cycle of $\bar \s_i$. These define $ \nu ( \bar \s_i ) $
and $ \nu ( \bar L ) = \{ \nu ( \bar \s_1 ) , 
\nu ( \bar \s_2 ) , \nu ( \bar \s_3 ) \}$. It will be convenient to define 
a group 
\bea 
 G ( \nu ( \bar L ) ) \equiv   G ( \nu ( \bar \s_1  ) ) 
\times G ( \nu ( \bar \s_2  ) )\times  G ( \nu ( \bar \s_3  ) )
\eea 
 $G ( \nu ( \bar \s_i  ) )$ is a product over  
cycle lengths of $\bar \s_i$.  If for a specified length, all 
the cycles have the same index, the group is defined to be the trivial 
group consisting of the identity permutation. If for a specified length, 
the cycles do not all have the same index, then the corresponding 
group is the group of all permutations of these cycles. As an example
say $\bar \s_1 $ is 
\bea 
&& \bar \s_1 = ( 1 ) ( 2 ) (3,4) (5,6) (7,8) (9,10,11) ( 12,13,14) \cr  
&& \nu ( \bar \s_1 ) = \{ 1 , 3, 1,3,1,1,1 \} 
\eea
Then $ G ( \bar \s_1)$ is $S_{2} \times S_3 $, where the $S_2$ 
permutes the cycles of length $1$ and the $S_3$ permutes the cycles of 
length $2$. There is no factor for the cycles of length $3$ since 
they have equal index. Another group which can permute cycles of 
the same length within each of the $ \bar \s_i$ is $\Aut (\bar L )$. 
We can define an intersection  $\Aut ( \bar L ) \cap G ( \nu ( \bar L ) )$
which consists of those permutations of like cycles in $\bar \sigma_i$
performed by  $G ( \nu ( \bar L ) )$ which can also be performed by  $\Aut ( \bar L )$.

We propose that the  general formula for refined counting is : 
\bea\label{genform}  
&& \cN ( T_1 , T_2 , T_3 ) =  \sum_{ l | d  }   { 1 \over l  { \bar d} ! }   
  \sum_{ [ \bar L ] } 
\hbox{ Number of labelled $\bar L $ in equiv class  $[ \bar L ]$ } \cr 
&& \qquad \qquad \qquad \times \hbox{ Number of epimorphisms }\times  | \Aut ( \bar L ) \cap G ( \nu ( \bar  L ) ) |  
\eea
For some of  the infinite sequences we consider  in Section 
\ref{sequence},  the simple form  (\ref{simpform}) suffices, for the one in section \ref{withones}  it is necessary to use the most general form (\ref{genform}). 
The factor $| \Aut ( \bar L ) \cap G ( \nu ( \bar  L ) ) |$, which we call the {\it index distribution 
factor}, has not been 
discussed explicitly in the mathematical literature, as far as we are aware. 
We illustrate simple  examples where it shows up in the appendix 
\ref{App:exampsNCF}.

Another perspective on the quotienting is given by the reverse construction 
where once starts from the labelled graph $ \bar L$, the indices $ \nu ( L ) $, 
and an epimorphism in order to reconstruct the covering ribbon graph $ L$.  
The reverse construction of going from labelled quotient graphs up
 to the covering graphs is described briefly in Appendix 
\ref{App:reverseconstruction}. It associates $Z_l$-valued 
{\it voltages} to the angles of the graph. We can also understand the weights, such as the  $ l^{ \bar d -1 } $ factor, leading to (\ref{genform})  
from the angle-voltage method.

\section{Some sequences of bi-partite graph counting  }  \label{sequence}

In this section we will illustrate the application of (\ref{genform}) 
to a variety of counting problems for ribbon graphs and bi-partite graphs. 
We will start with a simple example and then proceed to 
infinite sequences. 

\subsection{ $ \cN ( [3^4], [2^6], [6^2] ) $   } 

 We wish to know the number  $ \cN ( [3^4], [2^6], [6^2] ) $ of 
 ribbon graphs with four trivalent vertices
and two faces, each bounded by six edges. 
Equivalently, by graph duality,
we are considering ribbon graphs with two vertices with valency $6$ 
and four faces with valency $3$. In the language of bi-partite graphs 
we are considering the number of bi-partite graphs with four black vertices 
of valency $3$, six  bivalent white vertices. 

This is described by a triple of permutations $\s_1, \s_2, \s_3$ 
in $S_{12}$. Each triple is a labelled graph $L = \{ \s_1, \s_2, \s_3 \} $.
 The cycle structures  are 
\bea
[ \sigma   ]  \equiv \{ [ \s_1] , [ \s_2] , [ \s_3]  \} = \{[3^4], [2^6], [6^2] \}. 
\eea
The genus of the graph is $g=1$ as given by the Riemann-Hurwitz relation 
(\ref{RH}). Equivalence classes under conjugation 
(\ref{equivtriples}) correspond to ribbon graphs (also called unrooted maps
in the literature). Considering the 
possible divisors we have automorphisms generating $Z_l $ and corresponding
quotient graphs with $ \bar d = { d \over l } $ edges with 
\bea 
( l ,  \bar d ) = \{ ( 12 , 1 ) , ( 6,2 ) , ( 4,3) , ( 3,4) , ( 2,6) , (1,12) \} 
\eea 
The cycle structures of the quotient graph $ [\bar \sigma ] \equiv 
\{  [ \bar \s_1 ] ,  [ \bar \s_2 ] ,  [ \bar \s_3 ] \} $ are 
\bea
[ \bar \sigma  ]  = \{ [3^{s_{11}} 1^{s_{12}}], [2^{s_{21}} 1^{s_{22}}], [6^{s_{31}} 3^{s_{32}} 2^{s_{33}} 1^{s_{34}}] \} ,
\eea
where the integers $s_{ij}\ge 0 $ satisfy
\bea
3 s_{11} + s_{12} = 2 s_{21} + s_{22} = 6 s_{31} + 3 s_{32} + 2 s_{33} + s_{34} = \bar{d}. 
\eea
Applying the  Riemann-Hurwitz formula (\ref{RH}) the genus $g$ of 
the quotient graph is 
\bea
2g-2 &=& \bar{d} - (s_{11} + s_{12} + s_{21} + s_{22} + s_{31} +  s_{32} +  s_{33} + s_{34} ) \\ \nonumber \cr
  &=&  2s_{11} + s_{21} + 5s_{31} +  2s_{32} +  s_{33} - 2\bar{d} \, .
\eea
The branching  indices of the covering map of the Riemann surfaces  $ \Sigma ( L ) \rightarrow  \Sigma ( \bar L ) $ induced by 
the quotient $ L \rightarrow  \bar{L} $ are 
\bea 
\nu ( \bar L ) = [1^{s_{11}}3^{s_{12}}, 1^{s_{21}}2^{s_{22}}, 6^{s_{34}} 3^{s_{33}} 2^{s_{32}} 1^{s_{31}} ]
\eea 
We will sometimes omit the factors of $1$ in the index and simply write
\bea 
\nu ( \bar L ) = [3^{s_{12}}, 2^{s_{22}}, 6^{s_{34}} 3^{s_{33}} 2^{s_{32}} ]
\eea 
since they do not play any important role in the epimorphism equation.
To determine which $s_{ij}$ would contribute to the counting of $\cN (  [3^4], [2^6], [6^2]) $   we study the equations above for the indices 
and the related epimorphism equations. For instance the case $\bar{d} =1$, we get indices as $[2, 3, 6]$. However the epimorphism equation 
\bea 
e_1 e_2 e_3 = 1 ~~ ; ~~ e_1^2 = e_2^3 = e_3^6 =1 
\eea
has no solution for $e_i \in Z_{12}$. Note that the equations $e_i^{m_i} =1 $ 
mean that we have non-trivial monodromies which give $1$ after raising to 
the power $m_i$. Here they force $ e_1= \omega^{6} , e_2 = \omega^4 , e_3 = \omega^2$ for $\omega^{12} =1 $. The product gives $1$ but this is not an epimorphism, 
rather the image is the $Z_6$ subgroup generated by $\omega^2$. 

By the same analysis we find all possible contributions, which are listed in Table \ref{table1}. 
\begin{table}[ht] 
\caption{Possible quotient graph data for $[\sigma ] = \{ [3^4] , [2^6] , [6^2] \}$}
\centering 
\begin{tabular}{| c|  c |  c  |  c |} 
\hline  
$[ \bar \sigma ]$ & genus $g$ & $Z_l$  &  Indices $\nu [ \bar \sigma ]$ \\[0.5ex]
\hline 
$ \{[3^2], [2 1^4],[6] \}$ & $ 0 $ & $Z_2$ & $ \{ 1^2 ; 2^4;  1  \} ~ {\rm or }~   \{ ~  ; 2^4;  ~   \}$  \\
$\{ [3^2], [2^2 1^2], [3^2] \}$ & $0$ & $Z_2$ & $ \{ 1^2  ; 1^2 2^2 ; 2^2 \} ~ {\rm or } ~ 
 \{ ~   ; 2^2 ; 2^2 \}$  \\ 
$\{  [3^2], [2^3], [6] \} $ &   $1$ & $Z_2$ & $ \{ 1^2 ;1^3  ; 1 \}  ~{ \rm or }~   ~  \{ ~  ; ~  ; ~   \}$  \\  
$ \{ [3^4], [2^6], [6^2] \} $ & $1$ & $Z_{1}$ &  $\{ 1^4 ; 1^6  ; 1^2  \}  ~ { \rm or }~   \{ ~  ; ~  ; ~   \}$  \\ 
\hline
\end{tabular} 
\label{table1} 
\end{table}
For each quotient we analyse the epimorphism equation to determine 
the {\it epimorphism factors }, which are 
$1, 1, 3$, and $1$.  
In these examples the index distribution factor is $1$. 
The epimorphism equation  in the third case is 
\bea 
a_1 b_1 a_1^{-1} b_1^{-1} =1 
\eea
The solutions are $( a_1 , b_1 ) = \{ (1,\omega  ) , (\omega ,1) , (\omega ,\omega ) \} $ for $\omega^2 =1$.  The pair $( 1,1)$ solves the equation 
but, since there are no non-trivial indices, there is no epimorphism 
for this choice. In the typical case, where there are non-trivial 
indices, such solutions do contribute to a factor of $l^{2g}$ 
as we will see in the subsequent examples.

Finally we need the numbers of labelled graphs with these four cycle structures in Eq.~(\ref{table1}). They can be calculated by the characters of the related cycle structures, equation (\ref{charsumcz}), or SYMMETRICA on-line. Here we only list the results: $ 3 \times 5!   , 3 \times 5!, 5!$ and $3 \times 11!$. In summary, we obtain the number of the unrooted graphs by summing over all the contributions
\bea
\cN ( [3^4] , [2^6] , 6^2] ) =  
\left ( { 1 \times (3 \times 5!) \over 2 \times 6!}  + 
  { 1 \times (3 \times 5!) \over 2 \times 6! }  +
   { 3 \times 5! \over 2 \times 6! }  + 
{ 1 \times (3 \times 11!) \over { 1 \times 12! } }    \right ) = 1. 
\eea

\subsection{ $\cN (   [4p]  , [2^{ 2p }]  , [4p ]  ) $  } \label{chords} 

In this and following subsections, we will consider some 
infinite sequences of graphs.
 Here we consider the counting of the graphs described 
by permutation triples with cycle structures
 $ [\sigma ]  = \{ [4p] , [2^{2p}], [4p] \}$. Let us consider a general $Z_{l}$ quotient contribution from graphs with $[ \bar \sigma ]  = \{ [q], [2^{s_{21}} 1^{s_{22}}], [q] \}$, for all possible $l$ which divides $4p$ and we have $q = 4p/l$. The indices can easily be read off from the cycle structures $\sigma$ and $\bar{\sigma}$. We obtain
\bea
[ \nu ( \bar L )] = [l, 1^{s_{21}} 2^{s_{22}}, l]
\eea
or $[\nu ( \bar L )]  = [ l, 2^{s_{22}}, l]$ if we ignore the index $1$ as we will do in the following discussions. From the index structures we can conclude that when $l$ is odd $s_{22}$ must vanish to have a non-trivial result. That is because any index must be divisible by $l$ since the automorphism group is $Z_{l}$ for this case. The genus $g$ is calculated from Riemann-Hurwitz formula 
\bea
2g - 2 = q - (s_{21} + s_{22} + 1 + 1)
\eea
which leads to
\bea
g = { s_{21} \over 2}
\eea
which means $s_{21}$ must be an even integer. $s_{21}$ and $s_{22}$ satisfy a further constraint 
\bea \label{4pconstrain}
2 s_{21} + s_{22} =q
\eea
From this equation we deduce $s_{21} \leq \lfloor q/2 \rfloor $, where $\lfloor x \rfloor$ denotes integer part of $x$.

To count the number of the graphs applying our general formalism 
in Section \ref{QuotsAndRev}, we will calculate the number of labelled graphs
for  $[ \bar \sigma  ]  = \{ [q], [2^{s_{21}} 1^{s_{22}}], [q] \}$, and the corresponding {\it epimorphism factor}. The index distribution factor is $1$ in this case. 
Let us start with the {\it epimorphism factor}. From the indices $[l, 2^{s_{22}}, l]$ we obtain the following epimorphism equations,
\bea
\prod^{g}_{i=1} [a_i, b_i] \cdot f_1 f_2 \prod^{s_{22}}_{j=1} e_j = 1 \, , \quad e_j^{2} =  f_j^{l} = 1 \, , 
\quad {\rm with} \quad \{ e_i \, , f_i \} \in Z_l \,.
\eea
Since $e_j^{2} = 1$ we have $e_j = -1$ for all $j$, so the above equations simplify to 
\bea
\prod^{g}_{i=1} [a_i, b_i] = 1 \, , \quad  f_1 f_2  = (-1)^{s_{22}} \, , \quad  f_i^{l} = 1 \,. 
\eea
The number of solutions for the first equation is simply $l^{2g} = l^{s_{21}} $. The number of  solutions for the second equation is the number of positive integers $<l$ that are relatively prime to $l$, namely it is the Euler totient function $\phi(l)$. Putting these  two results together, we obtain the {\it epimorphism factor} for $[ \bar{\sigma} ] = \{ [q] , [2^{s_{21}} 1^{s_{22}}] , [q]\} $, 
\bea
{\rm {\it epimorphism \, factor}} = l^{s_{21}} \phi(l) \, .
\eea
We then consider the number of labelled graphs 
 $ [ \bar \sigma  ]  = \{ [q] , [2^{s_{21}} 1^{s_{22}}] , [q]\} $, using (\ref{charsumcz}).  
 For the case at hand, we exploit the fact that the character of 
 a permutation in conjugacy class $[q]$ in $S_q$ (i.e. having a single
 cycle of length $q$) is only non-zero for hook representations 
 with row lengths $[q-k,1^k]$. The dimension of such a hook 
representation is 
\bea 
  d_{[q-k, 1^k]} = {q-1 \choose k}
\eea
using the standard formula for dimensions. We  can write  
\bea  \label{4p-Zl}
\mathcal{Z}([q], [2^{s_{21}} 1^{s_{22}}] , [q]) 
&=& {|[q]|^2 |[2^{s_{21}} 1^{s_{22}}]| \over q!  } \sum_k { \chi_{[q-k, 1^k]} ([q])^2 
\chi_{[q-k, 1^k]} ([2^{s_{21}} 1^{s_{22}}]) \over d_{[q-k, 1^k]} }  \\ \nonumber \cr
&=& {(q-1)!^2 \over  s_{21}! \, s_{22}!  \, 2^{s_{21}} } \sum_{ k=0}^{q-1}  {q-1 \choose k}^{-1} \chi_{[q-k, 1^k]} ([2^{s_{21}} 1^{s_{22}}])
\eea
We used the fact that $ \chi_{[q-k, 1^k]} ([q]) = ( -1)^k$. 
The character appearing in the sum  can be written in terms of binomial coefficients and 
there are some remarkable identities obeyed by the sum of binomial coefficients  which imply
\bea 
\sum_{ k=0}^{ q-1} {\chi_{[q-k,1^k]  }  ([  2^{s_{21} } 1^{s_{22} } ] ) \over d_{[q-k, 1^k]} } =  { q   \over s_{21} +1 }  = { 2 s_{21} + s_{22} \over   s_{21} +1 } \, .
\eea 
The explicit formulae for the characters in terms of binomial coefficients 
are derived using the Murnaghan-Nakayama relations and are given in the appendix \ref{charactersApen}. 
It is important that $s_{21} $ is even here.

Putting  everything together we obtain the contribution $\{ [q], [2^{s_{21}} 1^{s_{22}}], [q] \}$ to the number of unrooted bipartite graphs with cycle structure $\{ [4p], [2^{2p}], [4p] \}$, 
\bea \label{unrooted4p}
\mathcal{N}_{Z_l}([q], [2^{s_{21}} 1^{s_{22}}], [q] )
= {\phi(l) \over 4p }
  { q  \choose 2 s_{21} } {l^{s_{21}} (2 s_{21})! \over  2^{s_{21}}  (1 + s_{21})! }  \, ,
\eea
where we have used the fact $s_{22} = q - 2s_{21}$. Summing over all the contributions for all possible $l$'s and $s_{ij}$'s with constraints, i.e. equation (\ref{4pconstrain}), we obtain the final result of the number of ribbon 
 graphs for $\{ [4p], [2^{2p}], [4p] \}$, 
\bea 
\mathcal{N}([4p], [2^{2p}], [4p])= { 1 \over 4 p }  \left [ \mu_p + \sum_{ ql = 4p } \phi ( l ) \sum_{ s_{21} =0 }^{ q/2 } { q  \choose 2s_{21} } \mu_{ s_{21}/2 } l^{ s_{21} } 
 + \sum_{ ql = 4p } \phi (l) l^{ q /2 } \mu_{ q/4} \right ] 
\eea
where the first sum is only over  $l$  even and second over   $l$ odd,  and $\mu_k$ is defined as
\bea 
\mu_k  = { (4 k)! \over 2^{2 k} ( 2 k +1)! }   \, .
\eea
We note our result agrees precisely with the counting of Chord diagrams~\cite{chord}. A transformation between chord diagrams and ribbon 
graphs is explained in Appendix \ref{appendix:chords}.
It is  similar to the transformation \cite{MR2} that relates  QED vacuum 
graph counting to ribbon graph counting. 

\subsection{ $  \cN  ( [4p],[2^{2p-1}1^2], [2^{2p}]  ) $   }\label{withones} 

Here we consider the graphs corresponding to the sequence $ [ \sigma  ]  = \{ [4p], [2^{2p-1} 1^2], [2^{2p}] \}$ which involve 1-cycles. 
It is easy to see the graphs have vanishing genus, $g = 0$, i.e  they are all planar. 
If we consider $Z_l$ quotients we get 
\bea 
[ \bar \sigma  ] = \{  [{ 4p \over l }] ,    [2^{s_{21}}1^{s_{22}}] ,   [2^{s_{31}} 1^{ s_{32} }]\} 
\eea
This leads to indices of the form  $[l, 2^{s }, 2^{s_{32}  }]$ where $s= s_{22} -1 $ 
or $s= s_{22} -2$.  Consequently we have the following epimorphism equation, 
\bea 
f  \prod^{s + s_{32} }_{i =1} e_i = 1 \, , \quad \quad e^2_i =1 \, , f^{l} =1 \, , \quad {\rm with} \quad
\{e_i \, , f  \} \in Z_l
\eea
This has  no  solution for $l >2$. The only possible quotients are the $Z_2$ 
quotient leading to 
\bea
[ \bar \sigma  ]  = \{ [2p], [2^{s_{21}}1^{s_{22}}], [2^{s_{31}} 1^{ s_{32} }] \}
\eea
and the quotient by the identity element gives 
\bea
[ \bar \sigma  ]  =\{ [4p], [2^{2p-1} 1^2], [2^{2p}] \}
\eea

In the case of the $Z_2$ quotient, the integers $s_{ij}\ge 0 $ satisfy the following conditions
\bea 
&& 2p - (1 + s_{21}+ s_{22}+ s_{31} + s_{32}) = 2g - 2 = -2 \, , \\ \nonumber \cr
&& 2 s_{21} + s_{22} = 2p \, , \qquad  2 s_{31} + s_{32} = 2p \, , \qquad  s_{22} > 0 \, ,
\eea
where $s_{22}$ cannot vanish because the conjugacy class $[2^{2p-1} 1^2]$ contains cycles of length one. 
From this we obtain,
\bea
s_{32} = 0 \, , \qquad  s_{22} = 2 \, ,
\eea

Let us pause here to analyse the indices  of 
$ [ \bar \sigma  ] =  \{ [2p], [2^{p-1} 1^{2}], [2^p] \}$. 
Since the indices of the 1-cycles cannot both be $2$, it follows that 
the indices are  $[2, 2^s]$, where $0 \leq s<2$. Simple analysis leads to the conclusion that
 only the epimorphism equation with $s = 1 $ has a solution, and the {\it epimorphism factor} is just $1$. The { \it index distribution factor} is $2$ for this case, since there are two possibilities of choosing which $1$ in $[2^{p-1} 1^{2}]$ of $\{ [2p], [2^{p-1} 1^{2}], [2^p] \}$ to have an index $2$. Thus we conclude that 
 $\{ [2p], [2^{p-1} 1^{2}], [2^p] \}$ will contribute with a factor of $2$. 
  
To finish we need to compute the number of labelled graphs, which again can be obtained from the corresponding characters,
\bea 
\mathcal{Z}([2p], [2^{p-1} 1^{2}], [2^p]) = { |[2p]| |[2^{p-1} 1^{2}]| |[2^p]|   \over (2p)!} 
\sum_R {\chi_{R}([2p]) \chi_{R}([2^{p-1} 1^2]) \chi_{R}([2^p])   \over d_R} \, ,
\eea
where the representation $R$ can only possibly be a hook, $R=[2p-k, 1^k]$. 
Plugging  in the characters and simplifying,  we obtain
\bea
\mathcal{Z}([2p], [2^{p-1} 1^{2}], [2^p]) =  (2p)! 2^{1 - 2p} \Big( \sum^{p-2}_{m =0}  
 {2m \choose m} { 2p-2m-2 \choose p-m-1 } 
 + { 2p-2 \choose p-1 }  \Big) \,. 
\eea 
The summation can be performed explicitly and leads to a very simple result,
\bea
\mathcal{Z}([2p], [2^{p-1} 1^{2}], [2^p]) = (2p)!/2 \, ,
\eea
and by the same logic we have
\bea
\mathcal{Z}([4p], [2^{2p-1} 1^{2}], [2^{2p}]) = (4p)!/2 \, .
\eea
Collecting everything we obtain the final result, the number of ribbon graphs described by permutations
with cycle structure $\{ [4p], [2^{2p-1} 1^2], [2^{2p}] \}$
\bea
\cN ( [4p], [2^{2p-1} 1^2], [2^{2p}] )  = {1 \over 4p} (2p + 2 \times p) = 1 \, .
\eea

\subsection{ $ \cN ([3^{2p}],[2^{3p}],[6p])$ } \label{section:6p}

Here we consider  trivalent ribbon graphs with $2p$ vertices described by
$ [ \sigma ]  = \{ [3^{2p}] ,[ 2^{3p} ] , [6p ] \}$. The genus for this case is $g=(p+1)/2$, which means $p$ must be an odd integer.
Consider a general $Z_l$ quotient, 
\bea
[ \bar \sigma  ]  = \{ [3^{s_{11}} 1^{s_{12}}], [2^{s_{21}} 1^{s_{22}}], [ { 6p \over l } ]   \} \, 
\eea
Define $q = 6p/l$. It is easy to see it has indices $[3^{s_{12}}, 2^{s_{22}}, l]$, which leads to following epimorphism equation,
\bea
\prod^g_{i=1} [a_i, b_i] \cdot \prod^{s_{12}}_{m=1} f_m \prod^{s_{22}}_{n=1} e_n \, h=1  \, , \quad 
f^3_m = e^2_n = h^l =1 \, , \quad {\rm with} \quad {f_i\, , e_i \, , h} \in Z_l \, .
\eea
From the conditions on $f$ and $e$ we see that the factor $(\prod^{s_{12}}_{m=1} f_m \prod^{s_{22}}_{n=1} e_n)$ can form an element of order at most $6$. 
Thus for the equation to have solutions $l$ can only be $1, 2, 3$, and $6$. We conclude here that beside the contribution from identity quotient, we also have contributions from $Z_2$, $Z_3$, and $Z_6$ quotients. We will consider each contribution separately. 

Let us start with the contribution of graphs with cycle structure 
$ [ \bar \sigma  ]  =  \{ [3^{2p}] , [2^{3p}]  , [6p ] \} $ from identity quotient, the number of the corresponding labelled graphs is given as
\bea \label{simp1}
\mathcal{Z}([3^{2p}] , [2^{3p}]  , [6p ])=
{|[3^{2p}]| |2^{3p}]| |[6p]| \over (6p)!} \sum_R { \chi_R([3^{2p}]) \chi_R([2^{3p}]) \chi_R([6p]) \over d_R }
\, .
\eea
Plug in the explicit results on the characters, see Appendix \ref{charactersApen}, we find this result can be simplified as
\bea 
\mathcal{Z}([3^{2p}] , [2^{3p}]  , [6p ])=
  {2(6p)! (3p -2 )! \over 12^{(p+1)/2} ({p+1 \over 2})! ({3p-3 \over 2})! } \, .
\eea
The identity is true only for $p$ being odd, which is the case for the problem at hand. Consequently we then obtain the contribution to the 
ribbon graphs with cycle structure $ \{ [3^{2p}] , [2^{3p}]  , [6p ] \} $ from the identity quotient
\bea \label{6p-Z1}
\mathcal{N}_{Z_{1}}( [3^{2p}] , [2^{3p}]  , [6p ] )
=\mathcal{Z}( [3^{2p}] , [2^{3p}]  , [6p ] )/(6p)! 
= {2 (3p -2 )! \over 12^{p+1 \over 2} ({p+1 \over 2})! ({3p-3 \over 2})! } \, .
\eea

Secondly let us consider the contribution from $Z_2$ quotient, i.e. 
\bea
[ \bar \sigma  ] = \{ [3^{s_{11}}1^{s_{12}}], [2^{s_{21}} 1^{s_{22}}], [3 p] \}\, .
\eea 
However since it comes from $Z_2$ quotient we cannot have index equal to $3$, so $s_{12}$ must vanish and consequently $s_{11} = p$, namely only 
$[ \bar \sigma ] = \{ [3^{p}], [2^{s_{21}} 1^{s_{22}}], [3 p] \}$ would contribute. The graphs have indices $[2^{s_{22}}, 2]$, from which we obtain the {\it epimorphism factor} for this case,
\bea
{\rm {\it epimorphism \, factor}}=2^g,
\eea 
where $g  = (s_{21} - p +1)/2$ is the genus. We note $s_{21}$ must be an even number for $g$ being an integer, since $p$ is required to be odd. Furthermore $s_{ij}$ satisfy the following constraints 
\bea \label{6psum1}
2{s_{21}}+ {s_{22}} = 3p \, , 
\eea
combine this with the fact $g \geq 0 $, we have $ p - 1 \leq s_{21} \leq \lfloor3 p/2 \rfloor$.

The number of the corresponding labelled graphs can again be represented in terms of characters,
\bea 
\mathcal{Z}([3^{p}], [2^{s_{21}} 1^{s_{22}}], [3 p]) =
{|[3^p]| |[2^{s_{21}} 1^{s_{22}}]| |[3p]| \over (3p)!} \sum_R { \chi_R([3^p]) \chi_R([2^{s_{21}} 1^{s_{22}}]) \chi_R([3p]) \over d_R } \, .
\label{3plabelled}
\eea
Use explicit form of the characters in Appendix \ref{charactersApen}, the above result and consequently the contribution to the counting of ribbon graphs with cycle structure
$ \{ [3^{2p}] , [2^{3p}]  , [6p ] \} $ can be greatly simplified as
\bea \label{6p-Z2}
\mathcal{N}_{Z_2}([3^{p}], [2^{s_{21}} 1^{s_{22}}], [3 p]) &= &
2^{2g}  \mathcal{Z} ( [ 3^p] , [ 2^{s_{21}} 1^{s_{22}}] , [3 p]) /((3p-1)! \, 6p) \\ \nonumber \cr
 &=&  {(6 g + 2 s_{22} - 5)! \over 3^g g! s_{22}! (3 g + s_{22} - 3)! } \, ,
\eea
where the factor $2^{2g}$ is the {\it epimorphism factor} for this case. 

Thirdly we have the contribution from the $Z_3$ quotient. 
For the same reason as that of $Z_2$ quotient, the possible graphs can only have cycle structure $\{ [3^{s_{11}} 1^{s_{12}}], [2^p], [2 p] \}$. Correspondingly it has index structure $[3^{s_{12}}, 3 ]$, and the epimorphism equation is given as
\bea
\prod^g_{i=1} [a_i, b_i] \cdot  \prod^{s_{12}+1}_{m=1} e_m =1  \, , \quad e_m^3=1 \, , 
\quad {\rm with} \quad e_m \in Z_3 \, ,
\eea
where $g = (2 s_{11} - p + 1)/2$ is the genus. It is easy to find the number of solutions of this epimorphism equation, which is given by \footnote{For a simple derivation of the number of solutions of this equation, see Appendix \ref{appendix:N_n}.}
\bea \label{equation:Nn}
2 \cdot 3^{2g-1 } (2^{s_{12}} - (-1)^{s_{12}}) \, .
\eea
For this case $s_{ij}$ satisfy following constraints, 
\bea \label{6psum2}
3 s_{11}+ {s_{12}} = 2p \, ,  \quad g =(2 s_{11} - p + 1)/2 \geq 0 \, ,
\eea
which give $ (p - 1)/2 \leq s_{11} \leq \lfloor2 p/3 \rfloor$.

Now let us determine the number of labelled graphs. As usual it can be written in terms of characters,
\bea
\mathcal{Z}([3^{s_{11}} 1^{s_{12}}], [2^p], [2 p]) 
= 
{|[ 3^{s_{11}} 1^{s_{12}}]| |[2^p]| |2p| \over (2p)! }\sum_R {\chi_R([3^{s_{11}} 1^{s_{12}}])\chi_R([2^p])\chi_R([2p]) \over d_R }    \, ,
\eea
where the representation $R$ is a hook, and the characters are given in Appendix \ref{charactersApen}. The result multiplied with the {\it epimorphism factor} $2 \times 3^{2g-1} (2^{s_{12}  } - (-1)^{s_{12}})$ can be simplified as
\bea \label{simp2} 
\mathcal{N}_{Z_3}([3^{s_{11}} 1^{s_{12}}], [2^p], [2 p]) 
&=& 2 \times {3^{2g-1 } (2^{s_{12}  } - (-1)^{s_{12}}) \over (6p)(2p-1)!} 
\mathcal{Z}([3^{s_{11}} 1^{s_{12}}], [2^p], [2 p]) \\ \nonumber \cr 
&= &
\sum^{s_{12}}_{t \geq 0, t=p+1 \, \, {\rm mod} \, 3} 
{ 3^{g-1} (p-2)! \over 4^{g-1} g! t! (s_{12} -t)!  ({p-3 \over 2})! } \, .
\eea
Note that  when $p=1, g=0$, by taking the limit carefully the right hand side goes to $1/3$ smoothly. 

Finally we have the contribution from the  $Z_6$ quotient, where $[ \bar{\sigma} ] = \{ [3^{s_{11}} 1^{s_{12}}], [2^{s_{21}} 1^{s_{22}}], [p] \}$ with genus $g = ( 2 s_{11} + s_{21} - p+1 ) / 2$. Then the constraints on $s_{ij}$ are given as, 
\bea \label{6psum3}
3 s_{11}+ {s_{12}} = p \, ,  \quad 2 s_{21}+ {s_{22}} = p \, , \quad  2 s_{11} + s_{21} - p + 1 \geq 0 \, ,
\eea
so we have $ \lfloor (p - s_{21})/2\rfloor \leq s_{11} \leq \lfloor p/3 \rfloor $, and 
$(p - 1) - 2 \lfloor p/3\rfloor \leq s_{21} \leq \lfloor p/2\rfloor $. 

The epimorphism equation from indices $[3^{s_{12}}, 2^{s_{22}}, 6]$ is given as
\bea
\prod^g_{i=1} [a_i, b_i] \cdot \prod^{s_{12}}_{m=1} f_m \prod^{s_{22}}_{n=1} e_n \, h=1 \, , 
\quad f^3_m =e^2_n =h^6=1 \, , \quad \{f_m, e_n, h \} \in Z_6 \, ,
\eea
where $g = (2 s_{11} + s_{21} - p + 1)/2$ is the genus. The number of solutions of
the epimorphism equation can be obtained similarly as equation (\ref{equation:Nn})
\bea
4 \times 6^{ 2g-1 }  (2^{s_{12} } - (-1)^{s_{12}}) \, , 
\eea
To fully determine the contribution from $Z_6$ quotient we need the number of labelled graphs
\bea
\mathcal{Z}([3^{s_{11}} 1^{s_{12}}], [2^{s_{21}} 1^{s_{22}}], [p]) 
=
{ |[3^{s_{11}} 1^{s_{12}}]| |[2^{s_{21}} 1^{s_{22}}]| |[p]| \over p! }\sum_R {\chi_R([3^{s_{11}} 1^{s_{12}}]) \chi_R([2^{s_{21}} 1^{s_{22}}]) \chi_R([p]) \over d_R }  \nonumber
\eea
where we sum over all the hook representations $R$. As previous cases, the number of labelled graphs multiplied with the {\it epimorphism factor} may be further simplified and we obtain the final contribution from $Z_6$ quotient,
\bea \label{simp3}
&& \mathcal{N}_{Z_6}([3^{s_{11}} 1^{s_{12}}], [2^{s_{21}} 1^{s_{22}}], [p]) \\  \nonumber \cr
 & = & 4 \times 6^{ 2g-1 }  (2^{s_{12} } - (-1)^{s_{12}}) \mathcal{Z}([3^{s_{11}} 1^{s_{12}}], [2^{s_{21}} 1^{s_{22}}], [p]) /(6\, p!)
\\  \nonumber \cr
&=&  
 \sum_{r,s,t \geq 0} 2{ 3^{ g-1 } ({p+r-4 \over 2})! \over r! s! t! g! ({p+r-6 \over 4} )! }  
\eea
with conditions on $s$, $t$ and $r$: 
\bea 
2s=p+2\,\, {\rm mod} \,\, 3, \quad 2t=p+1\,\, {\rm mod} \,\, 3 \, ,  
\eea 
and they further satisfy $(p + 6 - 3 r - 4 s - 4 t)/12 = g$. As in the  $Z_3$ quotient case, we can handle the special case by carefully taking the limit to  $p=1, g=0$, where  the right hand side of Eq.~(\ref{simp3}) goes to $1/3$. 
%, and $r, s, t$ are related to $s_{12}, s_{22}$ by
%\bea
%s + t = s_{12} + 3 (s_{22} - r)/4.
%\eea

The highly non-trivial simplifications  in  (\ref{simp1})(\ref{6p-Z2})(\ref{simp2})(\ref{simp3})
follow from comparing with the results of  \cite{triangulation}. Our computations 
are based on permutation group characters and involve alternating signs, whereas
the more direct combinatoric approach of \cite{triangulation} is a sum of positive numbers. 
We have checked these simplifications by inserting example $s_{ij}$  in Mathematica, 
but it is desirable to obtain first principles derivations of these simplifications of  
character sums since they could potentially be applied to more general $T_1, T_2, T_3$. 

We conclude here with the final result obtained by summing over all the contributions we have discussed, 
\newpage
\bea
&& \mathcal{N}([3^{2p}], [2^{3p}], [6p]) \\ \nonumber \cr  
&=& 
\mathcal{N}_{Z_1}([3^{2p}], [2^{3p}], [6p]) 
+
\sum_{s_{21}} \mathcal{N}_{Z_2}([ 3^p], [2^{s_{21}} 1^{s_{22}}], [3 p]) \\ \nonumber \cr 
&+&
\sum_{s_{11}} \mathcal{N}_{Z_3}([ 3^{s_{11}} 1^{s_{12}}], [2^p], [2 p])   
+ 
\sum_{s_{11}, s_{21}} \mathcal{N}_{Z_6}([3^{s_{11}} 1^{s_{12}}], [2^{s_{21}} 1^{s_{22}}], [p])
\eea
where the summation ranges are determined by equations (\ref{6psum1}), (\ref{6psum2}) and (\ref{6psum3}). Here we give a list of $\mathcal{N}([3^{2p}], [2^{3p}], [6p])$ for small $p=2l-1$. 
\begin{table}[ht]
\caption{Table of $\mathcal{N}([3^{2p}], [2^{3p}], [6p])$} 
\centering  
\begin{tabular}{c c c c} 
\hline\hline                        %inserts double horizontal lines
$l$ & $\mathcal{N}([3^{2p}], [2^{3p}], [6p])$ & $l$ & $\mathcal{N}([3^{2p}], [2^{3p}], [6p])$ \\ [0.5ex] % inserts table 
%heading
\hline                  % inserts single horizontal line
1 & 1 & 6 & 5849686966988  \\ % inserting body of the table
2 & 9 & 7 & 23808202021448662  \\
3 & 1726 & 8  & 136415042681045401661  \\
4 & 1349005 & 9 & 1047212810636411989605202 \\
5 & 2169056374 & 10 & 10378926166167927379808819918 \\ [1ex]      % [1ex] adds vertical space
\hline %inserts single line
\end{tabular}
\label{table:nonlin} % is used to refer this table in the text
\end{table}

\subsection{  $ \cN (  [3^k] , [3^k] , [3k ]  )  $ } 

We consider bi-partite graphs with $k$ black vertices of valency $3$, 
$k$ white vertices of valency $3$, and a single face. 
Graphs with general $Z_l$ quotient have cycle structures 
\bea
[ \bar \sigma ]  = \{ [3^{s_{11}}1^{s_{12}}], [3^{s_{21}}1^{s_{22}}], [3k/l] \},
\eea 
which implies the  following epimorphism equation, 
\bea
\prod^g_{i=1} [a_i, b_i] \cdot f \prod e_i =1, \quad f^l=e_i^3=1, \quad \{f, e_i \} \in Z_l
\eea
which only has solutions for $l=1$ and $l=3$. 
Namely only $[\bar \sigma  ] = \{ [3^k] , [3^k] , [3k ]  \}$ of identity quotient, and its $Z_3$ quotient 
$[\bar \sigma  ] = \{ [3^{s_{11}}1^{s_{12}}], [3^{s_{21}}1^{s_{22}}], [k] \}$ will contribute to the counting of 
ribbon graphs with cycle structure $\{ [3^k] , [3^k] , [3k ]  \} $. 

Let us start with the first case. The number of labelled graphs for $[\bar{\sigma}] = \{ [3^k] , [3^k] , [3k ]  \}$, 
\bea
\cZ([3^k] , [3^k] , [3k ]) &=& 
\sum_R { |[3^k]|^2  |3k| \over (3k)!} {\chi_{R} ([3^k])^2 \chi_R ([3k]) \over d_R} 
\\ \nonumber \cr
      &=& \sum^{k - 1}_{m=0} \sum^2_{q=0} {(-1)^{q+m}(3 m + q)! (3 k - 3 m -q- 1)!   \over (m! (k - 1 - m)!)^2 3^{2 k - 1} k } \\ \nonumber \cr
&=&
{  (1 + 3 l)! (1 + 2 l)! \over  (1 + l) l!^3 3^l } \, ,  
\eea
where at the last equality we have defined $k = 2l+1$ since the result is non-vanishing only for $k$ being an odd integer.

For $[ \bar \sigma ] = \{ [3^{s_{11}}1^{s_{12}}], [3^{s_{21}}1^{s_{22}}], [k] \}$ from the $Z_3$ quotient, $s_{ij}$ satisfies the following conditions
\bea
s_{i2} + 3 s_{i 1} = k \, ,  \quad 
2(s_{11} + s_{21}) - k -1 = 2g-2 \leq k-1 \,  .
\eea
Combining with the fact that genus $g = (s_{11} + s_{21}) - {k-1 \over 2} \geq 0$, we find the constrains on $s_{ij}$, 
\bea \label{3ksum}
s_{11}+s_{21} \geq {k-1 \over 2} \, , s_{11} \leq \lfloor {k \over 3} \rfloor \, , s_{21} \leq \lfloor {k \over 3} \rfloor \, .
\eea
So we have $(k - 1)/2 - s_{11} \leq s_{21} \leq  \lfloor k/3\rfloor $, and  $ (k - 1)/2 - \lfloor k/3\rfloor \leq s_{11} \leq \lfloor k/3\rfloor $.

The {\it epimorphism factor}  for $[ \bar \sigma ] = \{ [3^{s_{11}}1^{s_{12}}], [3^{s_{21}}1^{s_{22}}], [k] \}$ is determined by the number of the solutions of following epimorphism equation,
\bea
\prod^g_{i=1} [a_i, b_i] \cdot \prod^{s_{12}+s_{22}+2}_{m=1} f_m =1  \, , 
\eea
where the genus $g = s_{12} + s_{22} - (k-1)/2$, and $f^3_m =1$ from index $3$. So we obtain the number of the solutions of the monodromy equations
\bea
3^{2 g} (2^{s_{12}+s_{22}+2 } + 2(-1)^{s_{12}+s_{22}})/3. 
\eea
Finally the number of labelled graphs is given as,
\bea
\cZ([3^{s_{11}}1^{s_{12}}], [3^{s_{21}}1^{s_{22}}], [k]) 
&=& {|[3^{s_{11}}1^{s_{12}}]| |[3^{s_{21}}1^{s_{22}}]| |[k]| \over k!  }
\sum_R  { \chi_{R} ([3^{s_{11}}1^{s_{12}}]) \chi_R ([3^{s_{21}}1^{s_{22}}])  \chi_R ([k]) \over d_R} \, , \nonumber
\eea
where the representation is a hook, i.e. $R = [k - l, 1^l]$, and the characters can be found in Appendix \ref{charactersApen}. 

Now we are ready to write down the final result, 
\bea
\cN([3^k], [3^k], [3k]) &=& { 1 \over (3k)! }  \cZ([3^k], [3^k], [3k])    \\ \nonumber \cr
&+&   { 1 \over k! }  \sum_{s_{ij} } 3^{2 g - 2}  (2^{s_{12}+s_{22}+2 } + 2(-1)^{s_{12}+s_{22}}) \cZ([3^{s_{11}} 1^{s_{12}} ], [3^{s_{21}}  1^{s_{22}}], [k])
  \, , \nonumber
\eea
where the summation over $s_{ij}$ is according to Eq.~(\ref{3ksum}).
Just for illustration here we list a few examples of the result for small $k = 2l -1$,
\begin{table}[ht]
\caption{Number of $\mathcal{N}([3^k], [3^k], [3k])$} 
\centering  
\begin{tabular}{c c c c} 
\hline\hline                        %inserts double horizontal lines
$l$ & $\cN([3^k], [3^k], [3k])$ & $l$ & $\cN([3^k], [3^k], [3k])$ \\ [0.5ex] % inserts table 
%heading
\hline                  % inserts single horizontal line
1 & 1 & 6 & 10045237344  \\ % inserting body of the table
2 & 4 & 7 & 10197348792270  \\
3 & 190 & 8  & 14582208729414372  \\
4 & 37372 & 9 & 27949233397422911524 \\
5 & 14948524 & 10 & 69179942505290755525648 \\ [1ex]      % [1ex] adds vertical space
\hline %inserts single line
\end{tabular}
\label{table:nonlin1} % is used to refer this table in the text
\end{table} 

%%%%%%%%%%%%%%%%

It is interesting to compare the counting discussed here with the one in the previous 
subsection. Given a bi-partite graph with $ [\sigma] = \{ [ 3^k] , [3^k ] , [3k] \} $, 
we can convert all the black and white vertices to black vertices, leaving 
$2k$ black trivalent vertices and we introduce $3k$ white vertices in the middle of 
the edges. We now have a graph of type $[\sigma ] = \{ [3^{2k}]  , [2^{3k} ] , [6k] \} $ 
considered in section \ref{section:6p}. This procedure of doubling 
the edges is called ``cleaning''  in the context of Belyi theory (for a review and references on this see \cite{MR1}).  For example the 
$4$ bipartite graphs at genus $2$ from the second  row of table \ref{table:nonlin1}
give,  after cleaning, some of the $9$  in the second  row of table \ref{table:nonlin}. 
These $9$ are described in terms of polygon gluings in Figure 1 of  \cite{girondo}. 
The faces can  described by a standard permutation  $ ( 1, 2, \cdots , 18)$. 
The lines joining the labelled polygon edges define $\sigma_2$ in conjugacy class $[2^9]$. 
This allows us to work out $\sigma_1 = ( \s_2 \s_3)^{-1} $ which turns out to always have conjugacy class $[3^6]$ as expected.  Three of the $9$, denoted P2,P3,P6 in \cite{girondo} 
have the property that each cycle of their  $\s_1$ have all even or all odd numbers. 
This means that the vertices admit coloring by black and white so that all edges 
join black to white. On the other hand the four bi-partite graphs with $[\sigma ] = \{ [3^3] ,[3^3] ,[9]\} $ can easily be constructed using 
GAP, as described in \cite{MR2}. Applying the cleaning procedure at the level 
of the permutation triples, we find that two of the bipartite graphs become P2,P3 
respectively,  after cleaning. The remaining two  of them, which are not equivalent in $S_{9}$ 
become, after cleaning, equivalent in $S_{18}$.  It is worth noting that the interest in 
table \ref{table:nonlin1} from \cite{triangulation} is motivated by relations to extremal 
surfaces of hyperbolic geometry, and the $9$ from \cite{girondo} were worked out using 
classic results  of Fricke and Klein. For string theorists, interest in hyperbolic geometry is 
motivated by its connections to the moduli space of complex structures 
on punctured Riemann surfaces $\cM_{g,n}$. The extension of the present observations of the somewhat intricate relations between the $4$  and the $9$ to the complete sequences 
would be more than a mathematical curiosity for string theorists if it contains information about $\cM_{g,n}$. We leave this as an investigation   for the future.

\section{QFT counting with External Edges  }\label{sec:extedges}

Consider a quantum field theory counting problem
where we have $v_k$ vertices having  $k$  incident edges (with cyclic symmetry $Z_k$) and 
$E$ external legs. This can be described by $v_k$ black vertices 
of valency $k$.  By introducing 
a white vertex in  the middle of each internal line, we subdivide 
them into edges. Thus, in the resulting bi-partite graph, these white vertices have valency two.  
The extremities of the external edges are given white vertices (of valency one). 
Define $ d = \sum_k k v_k $.
There are  $ ( d -E   )  /2 \equiv M $ bi-valent  
white vertices  and $E$ univalent white vertices for the external legs.

In QFT counting problems,   we will have momenta 
and color generators associated with the external legs. These 
labelled external legs are apriori distinguishable. Of course 
in any given QFT, after doing the Feynman integrals we may find 
symmetries relating the amplitudes or Green's functions 
which differ by permutations of the external edges. In enumerating 
the Feynman integrals we have to do, we treat these external legs as distinguishable. 
It is thus useful to introduce a notion of  { \it external-edge-labelled } (EEL)  graphs appropriate 
for Feynman rules. Note that these are different from { \it labelled graphs} 
where all edges (not just external) are labelled and which are in 1-1 correspondence with
 permutation pairs (or triples). 

The subgroup $ S_E  \times S_{ 2M } $ of $S_d$ permutes the $E$ external edges 
and the $2M$ internal edges of the bi-partite graph. 
The symmetry of the internal white vertices is $ S_{M} [ Z_2 ]  $, generated 
by the $Z_2$ permutations of $M$ 2-cycles, along with the $S_M$ permutations 
of the cycles. Hence the following chain of subgroups will be useful  
\bea 
( S_E \times S_{M} [ Z_2 ] ) \rightarrow ( S_E  \times S_{ 2M } )   \rightarrow S_{ d }  
\eea

Using arguments of the kind employed in \cite{MR2}, we can see that these EEL 
graphs, with specified numbers $v_k$ of vertices of valency $k$ and $E$ of 
external edges, are in 1-1 correspondence with points of the double coset 
\bea 
X_{II} =  H_1 \setminus S_{ d  } / ( S_{M} [ Z_2 ] )
\eea
where $H_1 = \prod_k  S_{v_k } [ Z_k ]$.
This is closely related to another double coset  
\bea 
X_I =H_1  \setminus S_{d } / ( S_{M} [ Z_2 ] \times S_E ) 
\eea
$X_I$ describes unlabelled bipartite graphs with $k$ black vertices of 
valency $v_k$, $M$ white vertices of valency $2$ and with $E$ white vertices 
of valency $1$. One application in matrix model correlators of $X_I$ is 
in the computation of correlators where the holomorphic 
observable has $v_k$ traces of form $\prod_k ( tr Z^k)^{v_k} $ and the antiholomorphic observable 
has is $ ( tr ( Z^{\dagger})^2 )^M (tr Z^{\dagger} )^E$. 
 Points in  $X_I$, corresponding to connected graphs 
and refined according to face structure,  can be counted directly by the quotienting methods discussed in section \ref{QuotsAndRev}. It turns out, as we will explain, that the counting of points in $X_{II}$ corresponding to connected graphs 
and refined according to face structure, is in fact simpler after
 we make use of Hall's theorem.

Before we explain this point, let us elaborate on some general properties 
of $X_{I}$ and $X_{II}$ and a relation between the two of the form 
\bea\label{quotientrel}  
X_I = X_{II}/S_E 
\eea
$X_I$ is described in terms of pairs   
$ \sigma_1 , \sigma_2\in S_d$ in the specified conjugacy classes
\bea 
&&  \sigma_1 \in \prod_k [ k^{v_k} ] \cr 
&&  \sigma_2 \in [ 2^{M} , 1^E ]  
\eea
Note, in particular, that the 1-cycles of $\s_2$ can be any of $\{ 1 , \cdots , d \} $.  
Pairs related by a conjugation with $\gamma \in S_d  $ 
\bea 
\sigma_1' = \gamma \sigma_1 \gamma^{-1} \cr 
\sigma_2' =  \gamma \sigma_2 \gamma^{-1}
\eea
define the same point in $X_I$. An equivalent  description is to take 
$\gamma \in S_E \times S_{M } [S_2 ]  $ look at orbits 
of the action on 
\bea 
 \sigma_1 \in \prod_k [ k^{v_k} ]
\eea
To describe  $X_{ II} $,  we take 
\bea 
&&  \sigma_1 \in \prod_k [ k^{v_k} ] \cr 
&&  \sigma_2 \in [ 2^{M} ]   \qquad \sigma_2 \in S_{ 2M } \subset S_E \times S_{ 2M }  \subset S_{ d }  
\eea
If we regard $\s_2$ as a permutation in $S_{d}$, the one-cycles are fixed say 
 $ \{ (1) , (2) , \cdots , (E) \} $. This corresponds to fixing the labels
 of the external legs to  $ \{ 1 , 2 , \cdots , E \} $. In QFT we would label the 
momenta as $\{ p_1 , \cdots , p_E \} $. 
 The equivalence used to define $X_{II}$ is  
\bea
\sigma_1' = \gamma \sigma_1 \gamma^{-1} \cr 
\sigma_2' =  \gamma \sigma_2 \gamma^{-1}
\eea
with $ \gamma \in S_d $. It suffices to let  $ \gamma \in S_{ 2M }$. 
As in \cite{MR2}, 
 the description in terms of $\gamma \in S_d $ can be rephrased as a counting in terms of
orbits of $S_{2M} = {\rm Stabilizer\,\, of \,\,}  \bar \sigma_2  {\rm \,\, in\,\,}  S_d  
= S_{M } [ S_2 ] $ on $ \sigma_1 \in  \prod_k [ k^{v_k} ] $. 
But $S_M [ S_2] $ is in $S_{2M}$, so we can let $\gamma $ be in $S_{2M}$ 
to start with.

Points in   $X_{II} $  thus correspond to  
orbits of $S_M[ S_2]$ acting on the conjugacy class 
$\prod_k [k^{v_k}]$ in $S_d$. We take $ \sigma_1 , \sigma_1' \in \prod_k [ k^{v_k} ] $ 
to be in same orbit if 
\bea\label{SMS2orbact} 
\sigma_1' = \gamma \sigma_1 \gamma^{-1} 
\eea
with $ \gamma \in S_M[ S_2]$. 
This formulation  allows us to demonstrate an  action of $S_E$ on $X_{II}$.  
The action of $\mu \in S_{E} $ is to take 
\bea\label{SEorbact}  
\sigma_1 \rightarrow \mu \sigma_1 \mu^{-1} 
\eea
Since  the $S_E $ subgroup of  $S_{d} $ commutes with $S_{2M}$, 
the action of $S_E$ is well-defined on the equivalence classes
$X_{II}$.  Indeed, let 
\bea 
&& \mu \sigma_1 \mu^{-1} \equiv \tau_1 \cr 
&&  \mu \sigma_1' \mu^{-1} \equiv \tau_1'
\eea
We can show that  $\tau_1 , \tau_1' $ are conjugate in  $S_{M} [ S_2] $
\bea 
&& \tau_1' =  \mu \sigma_1' \mu^{-1} =  \mu \gamma \sigma_1 \gamma^{-1}  \mu^{-1} \cr
&& = \gamma \mu  \sigma_1  \mu^{-1} \gamma^{-1} = \gamma \tau_1 \gamma^{-1} 
\eea
If we mod out by this action of $S_E$, we get  $X_I$ which was defined 
as a quotient of $\sigma_1 \in T_1 $ by $ S_E \times S_{M}[Z_2]  $.  
This implies that the size of $X_{II} $ is bounded as : 
\bea\label{bound}  
|X_{II} | \le  E! | X_{I} |
\eea 
But we will soon obtain more precise information on the size of $X_{II}$.

\subsection{ Implications of Hall's theorem for counting 
ribbon graphs with labelled external edges } 

Let us now study the implications of Hall's theorem 
for the counting of ribbon graphs with external edges.

Now let's get to the computation of $ | X_{II}|$ which is more directly 
relevant in the Feynman graph context. Here we can again apply the 
Burnside Lemma to get a sum over $\gamma \in S_{M}[S_2] \subset  S_d $. 
This means that 
$\gamma$ leaves the $E$ elements fixed. But this means that $\gamma$ 
has 1-cycles. By Hall's theorem, all its cycle lengths are equal, so 
it must be the identity. So 
\bea \label{external}
&& | X_{II} | ={ 1 \over 2^M M! }  \sum_{ \sigma_1 \in T_1 }   
\sum_{ \sigma_3 \in T_3 } \delta ( \sigma_1 \bar \sigma_2 \sigma_3 )  \cr 
&& = { 1 \over (2M)! }  \sum_{ \sigma_i  \in T_i  } 
   \delta ( \sigma_1 \sigma_2 \sigma_3 ) \cr 
&& = { 1\over (2M)! } \hbox{ labelled graphs with vertex structures specified by 
$T_1,T_3$ and $E$ external edges }\cr 
&& 
\eea
The first wonderful consequence of this is that, as soon 
as we have one or more external legs, we can 
compute the desired numbers of Feynman graphs (i.e. EEL ribbon 
graphs) just as a sum over characters. 
So far we have used the Murnaghan-Nakayama Lemma to compute these, but it is worth noting 
that physical methods (free fermions) might also be useful \cite{MDfreeferm}.

Let us give a couple of examples of counting the number of EEL graphs.  As we saw in equation (\ref{external}) the counting of EEL 
graphs is fully determined by the number of corresponding labelled graphs, consequently they are just sums of characters. Thus 
we can simply take the results in previous section \ref{sequence} to obtain the number of EEL graphs. For instance let us 
consider the graphs with cycle structures $ [ \sigma ] =  \{ [2p], [2^{p-E} 1^{2E}], [2p] \}$. As we can read off from the cycle 
structures, the graphs have $2E$ external legs, one vertex and one face, while the genus is $g = (p-E)/2$. The number of 
labelled graphs of this type is readily read off from Eq.~(\ref{4p-Zl}), which we will quote here,
\bea
\mathcal{Z}([2p], [2^{p-E} 1^{2E}] , [2p])
=
(2p-1)!  { (2p)! \over  2^{p-E} (2E)!(p-E + 1)! } \, .
\eea
Dividing this result by $(2p-2E )!$, we obtain the number of EEL graphs of this type, 
\bea
\mathcal{N}_{X_{II}}([2p], [2^{p-E} 1^{2E}] , [2p])
=
 { (2p-1)! \over  2^{p-E} (p-E + 1)! } {2p \choose 2E}\, .
\eea

Another example we like to present is a trivalent interaction case, $[ \sigma ]  = \{ [3^{2p}], [2^{3p-E} 1^{2E}], [6p] \}$. It is also a higher genus case, the genus $g =(p-E+1)/2$. The graphs contain $2E$ external legs, $2p$ trivalent vertices and one face. And the number of EEL  graphs with this sort of cycle structure is already determined in equation (\ref{3plabelled}), i.e.
\bea
\mathcal{Z}([3^{2p}], [2^{3p-E} 1^{2E}], [6 p])/(6p-2E)! \, .
\eea

Suppose we are interested in $X_I$ to begin with, e.g for matrix model 
correlators. We know that 
\bea 
|X_I| && =  { 1 \over d! }  \sum_{ \gamma \in S_{d}  } 
\sum_{ \sigma_i \in T_i }  
                  \delta ( \sigma_1 \sigma_2 \sigma_3 ) 
\delta ( \gamma \sigma_1  \gamma^{-1} \sigma_1^{-1} ) 
 \delta ( \gamma \sigma_2  \gamma^{-1} \sigma_2^{-1} ) \cr 
&& =  { |T_2|  \over d! }  \sum_{ \gamma \in S_{d}  }
 \sum_{ \sigma_1 \in T_1 }   \sum_{ \sigma_3 \in T_3 } 
                  \delta ( \sigma_1 \hat  \sigma_2 \sigma_3 ) 
\delta ( \gamma \sigma_1  \gamma^{-1} \sigma_1^{-1} ) 
 \delta ( \gamma \hat  \sigma_2  \gamma^{-1} \hat \sigma_2^{-1} )\cr 
&& =  { 1  \over 2^M M! E!  }  \sum_{ \gamma \in S_{M}[ S_2] \times S_E    }
 \sum_{ \sigma_1 \in T_1 }   \sum_{ \sigma_3 \in T_3 } 
                  \delta ( \sigma_1 \hat \sigma_2 \sigma_3 ) 
\delta ( \gamma \sigma_1  \gamma^{-1} \sigma_1^{-1} ) 
\eea
In the second line, we have replaced the sum over all $\sigma_2$ 
having the integers $ \{ 1,  \cdots ,  E \} $ in 1-cycles 
and $ \{ E+1 , \cdots E + 2M \} $ in 2-cycles, with the size of this 
conjugacy class in $S_{2M}$ and a fixed $ \hat  \sigma_2$ in this conjugacy 
class in the permutation sums. 
The requirement to commute with this fixed  $ \hat \sigma_2 $, 
leads to the condition that $\gamma $ is in  $S_{M}[ S_2] \times S_E $
which is a subgroup of $ S_{2M} \times S_E $. 
From Hall's theorem, we know that $\gamma $ has cycles of equal length.  
In this case, we learn that the cycle length $l$ 
must be a divisor of $E $ and $2M$. In other words, it is a divisor of the 
$gcd( E , 2M )$. 
So the above sum can be written 
as 
\bea 
&& |X_I| =  { 1  \over 2^M M! E!  }  \sum_{ l | gcd( E , 2M ) } 
 { E! \over l^{ E/l} (E/l)! }  
N ( [l^{ 2M/l} ] ; S_{M} [ S_2 ] ) \times \cr 
&& ( \hbox{ number of }  \sigma_1 \hbox{ fixed by a 
 permutation of  type}  [l^{d/l} ] \hbox{ such that }  
\sigma_1 \hat \sigma_2 \hbox{  is in }  T_3 ) \cr 
&& ~~ 
\eea
The number $N ( [l^{ 2M/l} ] ; S_{M} [ S_2 ] )$ is the number of 
permutations with $2M/l$ cycles of length $l$ in the group $S_M [ S_2 ]$.
Using the cycle index of $S_M[S_2]$, it can be written 
more explicitly.  The multiplicity explained in words can be calculated using 
the quotienting method of section \ref{QuotsAndRev}.  
An immediate consequence of the above formulae is that 
when $ gcd ( 2M , E ) =1 $, then the sum over $l$ reduces to 
one term, with $l=1$. In other words the only automorphism
is the identity. In this case, we have 
\bea 
| X_{II} | = E! |X_I|
\eea

\section{ Geometry and topological field theory of  counting embedded graphs } 
\label{TFT} 

In this section we will revisit the Burnside formula 
for counting $ \cN ( T_1 , T_2 , T_3 )$. We will find that 
it has a geometrical  interpretation in terms of three dimensional 
topological field theory on a 3-manifold $X $  
with $S_d$ gauge group, a special case 
of the theories considered in \cite{DW}. It can also be interpreted 
in terms of covering spaces of $X$, which suggests 
an interpretation in terms of topological membranes.  We motivated the counting 
problem of $\cN ( T_1 , T_2 , T_3 )$ using the Hermitian matrix model and its string interpretation. The three dimensional interpretation here can be viewed 
as an uplifting of the strings covering a sphere $S^2$  to
membranes covering $S^2 \times S^1$.  The  simplified counting $ \cN ( T_1 , T_2  )$ has an interpretation in two-dimensional topological field theory. 
We expect that this connection between refinement 
and dimensional uplifting should be a  general theme.

We start from the expression of the counting in terms of $ \sigma_1 , \sigma_2, \s_3 $ 
which are elements of conjugacy classes $ T_1, T_2 , T_3 $ of $S_d= G $. A concrete example to 
bear in mind (from Section \ref{sequence}) is  ribbon graph counting 
 with $2p$ cubic vertices so that  $T_1 = [3^{2p}]$ 
is the conjugacy class of permutations with $2p$ cycles of length $3$ 
and $T_2 = [2^{3p}]$ is a conjugacy class of pairings i.e. $3p$ cycles 
of length $2$. The conjugacy class $T_3$ keeps track of the faces of 
the embedded graph. The permutations act on 
$ \{ 1 , 2 , \cdots , 6p \equiv d \}$.

 Using the Burnside Lemma, the counting of these bi-partite graphs
 is given by 
\be\label{deltacosets} 
\cN ( T_1 , T_2 , T_3 ) = 
{1 \over |G|} \sum_{\gamma\in S_d} \sum_{ \sigma_i \in T_i  }
  \delta(\gamma \sigma_1 \gamma^{-1} \sigma_1^{-1} ) \delta(\gamma \sigma_2 \gamma^{-1} \sigma_2^{-1} ) 
\delta( \sigma_1 \sigma_2 \sigma_3 )
\ee
In the simpler counting problem but only keep track 
of $T_1, T_2$ we get 
\bea 
\cN ( T_1 , T_2 ) 
 = {1 \over |G|} \sum_{\gamma\in S_d} \sum_{ \sigma_1 \in T_1   } \sum_{ \sigma_2 \in T_2   }
  \delta(\gamma \sigma_1 \gamma^{-1} \sigma_1^{-1} ) 
\delta(\gamma \sigma_2 \gamma^{-1} \sigma_2^{-1} ) 
\eea

We will exploit the fact that these 
conjugacy classes  can be identified with representatives of cosets 
$ G/H_1, G/H_2 , G/H_3 $, with $G= S_d$ and $H_i$ 
being subgroups which commute with a fixed element in the conjugacy classes. 
For any $ ( \sigma_1 , \sigma_2 ) $ we can write 
\bea 
\sigma_1 = \alpha_1  \hat \sigma_1   \alpha_{1}^{-1}  \cr 
\sigma_2 =  \alpha_2  \hat \sigma_2   \alpha_{2 }^{-1} \cr 
\sigma_3 =  \alpha_3  \hat \sigma_3  \alpha_{3 }^{-1}
\eea
for some $\alpha_1, \alpha_2, \alpha_3$, 
where $  \hat \sigma_1 ,  \hat \sigma_2  , \hat \sigma_3 $ 
are some chosen representatives of the conjugacy class. In the example
we can take  $ \hat \s_1 =   (1,2,3) \cdots  ( 6p-2 , 6p-1, 6p ) $, 
$ \hat \s_2 =  ( 1,2) ( 3,4) \cdots ( 6p-1 , 6p )$ and $\hat \s_3 = ( 1, 2, \cdots , 6p ) $. The $H_i$  subgroups 
are generated by cyclic permutations for each cycle of $\hat \s_i$ 
and permutations which exchange the cycles. They are wreath products
with  $ H_1 = S_{2p} [ Z_3] $,    $H_2 = S_{3p} [ Z_2]$, $H_3= Z_{6p}$.
We can convert the sums over $\s_i$ into sums 
over $ \alpha_i$, by introducing factors ${ 1 \over |H_i| } $ required 
since the multiplication of $ \alpha_i$ on the right by $H_i$ does not change 
$\s_i$. 
\bea 
\cN ( T_1 , T_2 , T_3 ) = 
&& {1 \over |G| |H_1| |H_2 | |H_3| } 
 \sum_{ \alpha_i \in G } \sum_{ \gamma \in S_d } 
  \delta(\gamma \alpha_1 \hat \sigma_1 \alpha_1^{-1} 
  \gamma^{-1}    \alpha_1 \hat \sigma_1^{-1}  \alpha_1^{-1}     )
 \delta(\gamma   \alpha_2 \hat \sigma_2 \alpha_2^{-1} 
\gamma^{-1}    \alpha_2 \hat \sigma_2^{-1}  \alpha_2^{-1}     ) \cr 
&& \qquad \qquad \qquad \qquad \delta( \alpha_1 \hat \sigma_1 \alpha_1^{-1}  \alpha_2 \hat \sigma_2
 \alpha_2^{-1}  \alpha_3 \hat \sigma_3 \alpha_3^{-1}  )
\eea 
The equations 
\bea\label{genuseq} 
&& \gamma \alpha_1 \hat \sigma_1  \alpha_1^{-1}  \gamma^{-1} = 
 \alpha_1  \hat \sigma_1 \alpha_1^{-1}  \cr 
&&  \gamma \alpha_2 \hat \sigma_2 \alpha_2^{-1} \gamma^{-1} =  \alpha_2 
\hat \sigma_2 \alpha_2^{-1} 
\eea
from (\ref{deltacosets})  imply that $\alpha^{-1}_1 \gamma \alpha_1 \in H_1 $ 
 and $\alpha^{-1}_2 \gamma \alpha_2 \in H_2$. 
So we can convert the  sums over conjugacy classes $ \sigma_1 , \sigma_2  $
 into sums over $H_1 , H_2 $. 
\bea\label{countsymmconj} 
&&  \cN ( T_1 , T_2 , T_3 ) \cr 
&& =  {1 \over |G||H_1||H_2| |H_3| } 
\sum_{ \alpha_1 , \alpha_2 , \alpha_3 \in G } 
\sum_{ u_1  \in H_1 } \sum_{ u_2 \in H_2 } 
\delta ( \alpha^{-1}_1 \gamma \alpha_1 u_1 ) 
\delta ( \alpha^{-1}_2 \gamma \alpha_2 u_2) 
\delta (\alpha_1 \hat \sigma_1 \alpha_1^{-1}  ~ \alpha_2 \hat \sigma_2 
\alpha_2^{-1} ~ \alpha_3 \hat \sigma_3 \alpha_3^{-1}  )  \cr 
&& = {1 \over |G||H_1||H_2| |H_3| }  \sum_{ \alpha_i \in G } 
 \sum_{ u_1  \in H_1 } \sum_{ u_1 \in H_2 } \delta ( \alpha_1 u_1  
\alpha_1^{-1} \alpha_2 u_2 \alpha_2^{-1} )
\delta (  \hat \sigma_1 \alpha_1^{-1}  \alpha_2 \hat \sigma_2 
\alpha_2^{-1}  \alpha_3 \hat \sigma_3 \alpha_3^{-1} 
 \alpha_1  ) \cr 
&&  = {1 \over   |G| |H_1||H_2| |H_3|  }  \sum_{ \alpha_{12} ,  \alpha_{23} , 
\alpha_{31} \in G } 
\sum_{ u_1  \in H_1 } \sum_{ u_2 \in H_2 } 
\delta ( \alpha_{21}  u_1 \alpha_{12} u_2 ) 
\delta (    \hat \sigma_1 \alpha_{12}  \hat \sigma_2 \alpha_{23} 
   \hat \sigma_3 \alpha_{31}  ) \delta ( \alpha_{12} \alpha_{23} \alpha_{31} ) 
\eea 
We defined $ \alpha_{ij} = \alpha_i^{-1} \alpha_j $ which implies 
$ \alpha_{ij} = \alpha_{ji}^{-1} $.  

The simpler counting function $ \cN ( T_1 , T_2 ) $ can be written with these 
steps as 
\bea 
\cN ( T_1 , T_2 ) = 
{ 1 \over |H_1| |H_2| } 
\sum_{\alpha_{12} \in G } \sum_{ u_1 \in H_1 } \sum_{ u_2 \in H_2 } 
\delta ( u_1 \alpha_{12} u_2  \alpha_{12}^{-1} )
\eea
This formula can be used to express the counting in terms of cycle 
indices of $H_1 , H_2$ \cite{MR2}.  
The expression (\ref{countsymmconj}) is more complicated 
in that it involves constraints linking $u_i \in H_i $ 
and expressed in terms of the $\hat \s_i$. 
We have exploited the Burnside 
formula along with quotienting methods to explicitly count it, 
but here we will continue our focus on the geometry of the formulae. 
The expression (\ref{countsymmconj}) is nicely symmetric 
under $T_1,T_2,T_3$ permutations. 
We can also write a shorter less symmetric expression, by solving the 
last delta function. 
\bea 
\cN ( T_1 , T_2 , T_3 ) 
&& =  {1 \over   |H_1||H_2| |H_3|  }  \sum_{ \alpha_{12} ,   
\alpha_{31} \in G } 
\sum_{ u_1  \in H_1 } \sum_{ u_2 \in H_2 } 
\delta ( \alpha_{12}^{-1}   u_1 \alpha_{12} u_2 ) 
\delta (    \hat \sigma_1 \alpha_{12}  \hat \sigma_2 \alpha_{12}^{-1}  
   \alpha_{31}^{-1} \hat \sigma_3 \alpha_{31}  ) \cr 
&& =  {1 \over   |H_1||H_2|  }  \sum_{ \alpha_{12} \in G } 
\sum_{ u_1  \in H_1 } \sum_{ u_2 \in H_2 } \sum_{ \sigma_3 \in T_3 } 
\delta ( \alpha_{12}^{-1}   u_1 \alpha_{12} u_2 ) 
\delta (    \hat \sigma_1 \alpha_{12}  \hat \sigma_2 \alpha_{12}^{-1}  
   \sigma_3   )
\eea
The sums over $H_1, H_2$ can be turned into sums over $G$
by introducing extra delta functions. 
\bea 
&& \cN ( T_1 , T_2 , T_3 ) =  {1 \over   |H_1||H_2|  }  \ \sum_{ \alpha_{12} \in G } 
\sum_{ u_1 , u_2  \in G  }  \sum_{ \sigma_3 \in T_3 } 
\delta ( \alpha_{12}^{-1}   u_1 \alpha_{12} u_2 ) 
\delta (    \hat \sigma_1 \alpha_{12}  \hat \sigma_2 \alpha_{12}^{-1}  
   \sigma_3   )\cr 
&& \qquad \qquad \qquad \qquad \delta ( u_1 \hat \sigma_1 u_1^{-1}  \hat \sigma_1)
  \delta ( u_2 \hat \sigma_2 u_2^{-1}  \hat \sigma_2)
\eea

%
%
%%%%%%%%%%%%%%%%%%
\begin{figure}[h]
\scalebox{1.0}{
\centerline{\includegraphics[height=4.5cm]{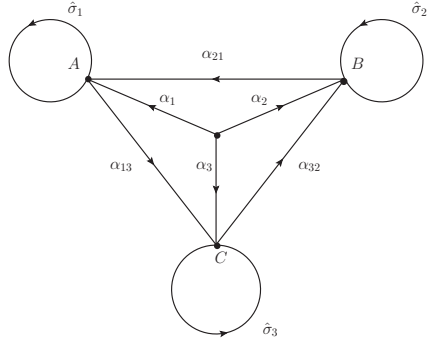} }
%\centerline{\includegraphics{triangle} } 
}
\caption{\it
Paths on the sphere  }
 \label{fig:alphas-picture}
 \end{figure}
%%%%%%%%%%%%%%%%%%%
%
%
 We can also work with a maximally  symmetric expression (\ref{countsymmconj})
where we have : 
\bea\label{countsymmconjI}
&& \cN ( T_1 , T_2 , T_3 ) \cr 
&& =  {1 \over   |H_1||H_2| |H_3|  }  \sum_{ \alpha_{12} ,  \alpha_{23} , 
\alpha_{31} \in G } 
\sum_{ u_1  \in H_1 } \sum_{ u_2 \in H_2 }   \sum_{ u_3 \in H_3 }  
\delta (    \hat \sigma_1 \alpha_{12}  \hat \sigma_2 \alpha_{23} 
   \hat \sigma_3 \alpha_{31}  ) \delta ( \alpha_{12} \alpha_{23} \alpha_{31} ) 
 \cr 
&& \qquad \qquad \qquad \qquad \qquad \delta ( u_1 \hat \sigma_1 u_1^{-1} \hat \sigma_1^{-1} )  
   \delta ( u_2 \hat \sigma_2 u_2^{-1} \hat \sigma_2^{-1} )
 \delta ( u_3 \hat \sigma_3 u_3^{-1} \hat \sigma_3^{-1} )\cr 
&& \delta ( \alpha_{12}^{-1}   u_1 \alpha_{12} u_2 ) \delta ( \alpha_{13}^{-1}   u_1 \alpha_{13} u_3 ) \delta ( \alpha_{23}^{-1}   u_2 \alpha_{23} u_3 ) 
\eea
In the last line we have introduced extra delta functions, implied by the existing ones, to make everything look symmetric.

To understand the manipulations associated with the application 
of Burnside theorem geometrically, it is useful to consider 
a topological space $X$ whose fundamental group can be described 
in terms  of generators and relations of the kind we encounter 
in the Burnside manipulations. It is useful to recall that 
the fundamental group of a cell complex can be obtained 
by choosing a cell decomposition, considering the group
generated by sequences of 1-cells forming closed paths and 
relations coming from 2-cells.  Given the counting formulae we have written 
we can ask what type of topological space has one-cells corresponding 
to the group variables we are summing and 2-cells for the delta functions 
that appear in the sums. 

Then the sums can be interpreted as counting 
homomorphisms from the fundamental group of $X$ to $S_d$. It is known 
that these homomorphisms also count covering spaces of $X$. This line of 
argument was used in developing the string interpretation of the large 
$N$ expansion of 2D Yang-Mills theory. Here we will follow the same logic 
and will be lead to 3-manifolds related to the counting of 
ribbon graphs and  bi-partite graphs. We will discuss the physical 
interpretation of the 3-manifold and its covers.  Since our interest 
is in homomorphisms from $\pi_1$ to $S_d$, we will use the same notation 
for paths as for their corresponding permutations.

The starting point is to consider the equation $\s_1 \s_2 \s_3 =1$. 
As recalled in Section 2, this corresponds to counting branched covers of 
the sphere with three branch points, equivalently unbranched covers of 
sphere minus three discs. The fundamental group of the sphere minus three points is generated by three elements multiplying to $1$. 
A  geometrical picture of the 
$\alpha_i , \hat \s_i  , \alpha_{ij} $ variables is given by Figure 
\ref{fig:alphas-picture}. We think of sphere as plane with infinity identified to a point and $ \hat \s_i$ surround the three discs that have been removed. 
A central basepoint is joined to three base-points on the disc boundaries by 
$\alpha_i$. 

\begin{figure}[h]
\scalebox{1.0}{
\centerline{\includegraphics[height=4.5cm]{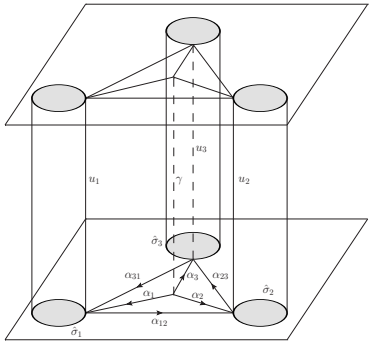} }
%\centerline{\includegraphics{triangle} } 
}
\caption{\it
The 3-manifold and the paths corresponding to permutations used in Burnside formula for ribbon graphs  }
 \label{fig:s2s1paths}
 \end{figure}

This picture does not show the $ u_i$  or $\gamma $ permutations appearing in 
the equations of this section, nor their relations with the 
$\alpha_i , \sigma_i$. In order to get the more complete picture, 
the appropriate space to consider is a 3-manifold  
$X \equiv  S^2 \setminus ( D_2 \cup D_2 \cup D_2  ) \times S^1$.
This is the same as $( S^2 \times S^1 ) \setminus ( \tilde T_2 \cup \tilde T_2 
\cup \tilde T_2  ) $  where $ \tilde T_2$ is the solid torus. 
To visualize this, we continue to think of  $S^2$ as $R^2$ with infinity 
identified to a point, thus drawing  $S^2$ as a planar sheet. 
The $S^1$ can constructed as an interval with the two points 
at the ends identified. The geometry $X$ is constructed by 
starting with $S^2$ at the bottom of the Figure \ref{fig:s2s1paths}, 
taking out the three discs, and evolving the geometry vertically 
and then identifying the $S^2 \setminus  ( D_2 \cup D_2 \cup D_2  )$ 
at the top with the one at the bottom. 
In the Figure \ref{fig:s2s1paths} we have also displayed 
the paths corresponding to the permutations $ \gamma , u_i$  that appear in 
our discussion of the Burnside Lemma applied to 
the refined counting of ribbon graphs. All the relations encountered
in the previous equations follow from 2-cells in the discretization of $X$
described in the figure.

%
%
%%%%%%%%%%%%%%%%%%
%%%%%%%%%%%%%%%%%%%
%
%

\begin{figure}[h]
\scalebox{1.0}{
\centerline{\includegraphics[height=4.5cm]{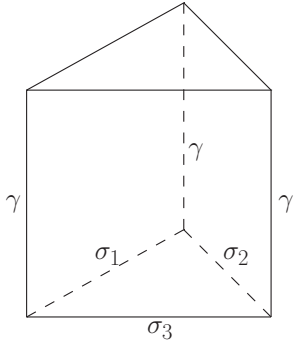} }
%\centerline{\includegraphics{triangle} } 
}
\caption{\it
A simpler cell decomposition}
 \label{fig:simple-cell-decomp}
 \end{figure}

The fundamental group of $X$ has  three generators 
$\sigma_1 , \sigma_2 , \sigma_3$ corresponding to paths 
starting from a central base-point  and going round each of the three discs ; 
along with a loop winding round the $S^1$ which we will call $\gamma$. 
There are relations among these generators 
\bea 
&& \sigma_1 \sigma_2 \sigma_3 = 1 \cr 
&& \gamma \sigma_1 \gamma^{-1} = \s_1 \cr 
&& \gamma \sigma_2 \gamma^{-1} = \s_2 \cr 
&& \gamma \sigma_3 \gamma^{-1} = \s_3 
\eea
A simpler cell decomposition of $X$ exists
(see Figure \ref{fig:simple-cell-decomp})   which contains just the 1-cells 
associated with permutations appearing in the simplest formula
 we started with (\ref{deltacosets}). The three vertical edges are identified and  there is also the 
identification of the lower triangle with the upper.  This description of 
$X$ was used in the context of 3D topological field theory 
in \cite{DW}. 

\noindent 
{\bf Topological field theory } 

The formula for $\cN ( T_1 , T_2 , T_3 ) $ can be interpreted 
as a partition function for topological field theory $ X = ( S^2 \setminus(  D_2 \cup D_2 \cup D_2 ) ) \times S^1$  of the kind considered by 
\cite{DW}. We  have a partition function which is lattice gauge theory 
for a finite group, which we will take to be $S_d$. There are group 
variables on the links. There is a product of weights, one for 
each 2-cell, which is just the delta function for the product of group 
elements around the 2-cell. This is topologically invariant, i.e
gives the same answer under refinements of the cell decomposition.
It is a special case of \cite{DW} where the 3-cocycles are chosen to 
be trivial. They have a more general topological action where 
3-cocycles give weights for each 3-cell. In our case this weight is just 1. 

In the special case, where $T_1, T_2 , T_3$ are all equal 
to the conjugacy class of the identity permutation,
$\cN ( T_1 , T_2 , T_3 ) $ is equal to $1$, which is the 
partition function for $S^2 \times S^1$ given in \cite{DW}. 
Thus a combinatoric interpretation of the partition function 
of $S^2 \times S^1$ for this simplest topological field theory with $S_d$ gauge group is given in terms of ribbon graph counting. It is interesting to ask 
if more general TFT questions with $S_d$ gauge group, 
e.g with non-trivial cocycles have a connection to ribbon graphs.

The topological theory is just counting homomorphisms 
from $\pi_1 ( X ) $ to $S_d$. This is the same as counting 
covers of $X$, weighted by the inverse of the order of the automorphism group
of the cover.  This can be viewed as topological membrane theory 
for membranes  wrapping  $X$.  The logic used to arrive at this conclusion 
is the same as  in the connection between large $N$ 2d Yang Mills 
and strings (see \cite{grta,cmr}), the key ingredients being 
delta functions over symmetric groups, the connection between covering 
spaces and homomorphisms between $\pi_1 (X) $ and $S_d$. 
It would be interesting to explore analogies between this combinatoric/topological version of string/membrane connection with the physical one linking 
Type IIA string theory in 10 dimensions and membranes in eleven dimensions
\cite{wittM}. For discussions of topological membranes in M-theory 
see  \cite{ams,bbcn,harmoore,dijkvaf}. 

%
%
%%%%%%%%%%%%%%%%%%
%%%%%%%%%%%%%%%%%%%
%
%

\noindent
{ \bf Back to the unrefined counting }. 

\begin{figure}[h]
\scalebox{1.0}{
\centerline{\includegraphics[height=4.5cm]{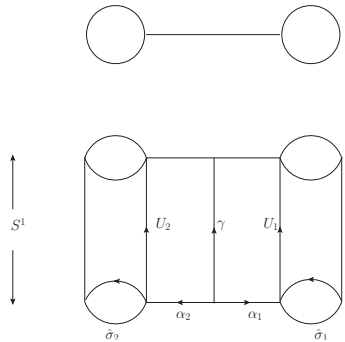} }
%\centerline{\includegraphics{triangle} } 
}
\caption{\it
Graph times $S^1$ explains the counting formulae with $T_1,T_2$ fixed but product unrestricted}
 \label{fig:graphs1}
 \end{figure}

We had seen a torus and a cylinder 
description of the counting in \cite{MR2}. Here we have constrained 
$\sigma_3 \in T_3$ and found a 3D geometry. How can we understand 
the unrefined case from this point of view ?  When we drop  the 
constraint on the product, we have two free generators. The graph 
with one vertex and two loops has a fundamental group 
generated by two elements. This can be viewed as obtained 
 from $S^2 \setminus  ( D_2 \cup D_2 \cup D_2 ) $ by dropping one disc
as well as the 2-cell. The application of Burnside Lemma introduces 
the extra circle. So we have the two-petal flower graph 
times $S^1$ as in Figure \ref{fig:graphs1}.    

Given our earlier discussion, this unrefined counting is a special case of 
the 3D problem of counting covers of $ (S^2\setminus ( D_2 \cup D_2 \cup D_2  ) ) 
 \times S^1$, with $T_1, T_2$ determining the windings around 
two of the boundaries and the third $T_3$ being summed over all partitions 
with equal weight.

%
%
%%%%%%%%%%%%%%%%%%
%%%%%%%%%%%%%%%%%%%
%
%

 For the unrefined counting of unlabelled graphs, 
$T_3$ unspecified, it suffices to have a 2D picture for the equations. 
For the refined counting, we need a 3D picture. 
The connection between refinement of counting and 
dimensional uplift, conversely between coarsening of counting 
and dimensional reduction should be a fairly general story.  
Refinements of counting often have a flavour of categorification.
The connection between categorificatrion and dimensional uplift has 
 been discussed in  \cite{benzviTFTITP}. 

\noindent
{\bf  A generalized counting to interpolate between labelled and unlabelled }  

The counting of $ \cZ ( T_1 , T_2 , T_3 ) $  comes up in 2dYM. 
We have related the counting of $ \cN ( T_1 , T_2 , T_3 ) $ 
to  $ \cZ  ( S^2 \times S^1 ; T_1  , T_2 , T_3 )$. 
We can generalize this by including a coupling constant $\lambda$ as
$  \cZ ( S^2 \times S^1 ; \hat \s_1 , \hat \s_2 , \hat \s_2 , \lambda  ) $
\bea 
 && \cZ ( S^2 \times S^1 ; T_1 , T_2, T_3 ,  \lambda ) = 
\sum_{\sigma_i \in T_i  } \sum_{ \gamma \in S_n } 
 e^{ - \lambda  ( C_{\gamma }   - d ) } \delta ( \g \sigma_1 \g^{-1} \sigma_1^{-1} ) 
   \delta ( \g \sigma_2 \g^{-1} \sigma_2^{-1} ) 
\delta ( \sigma_1 \sigma_2 \sigma_3 )\cr 
&&
\eea 
In the limit where $\lambda $ goes to infinity, the number of cycles 
$C_{\gamma}$ in the permutation $\gamma \in S_d$ must be $d$, which 
means $\gamma $ is the identity permutation. In the limit where $g$ goes to 
$0$, all cycle structures of $\gamma $ are summed with equal weights so we 
get $ \cN ( T_1 , T_2 , T_3)$. Explicit formulae for any $g$ can be written 
down for the examples considered in \ref{sequence} using the 
formulae given there.

\section{ Discussion} 

 The computation of correlators or amplitudes in quantum field 
theory rely, quite universally, on the use of an element of graphical 
enumeration  along with the computation of associated integrals. 
That is certainly the case for standard Feynman rules
in scalar field theory, or QED or the standard model. It is also 
true for the large $N$ expansions where scalar fields, gauge fields 
or fermions are promoted to matrices of a size $N$ that is treated
as a parameter. In the MHV-rules approach to amplitudes, inspired 
by twistor string theory,  
 amplitudes are constructed by gluing MHV vertices. 
The combinatoric aspect of finding different ways of gluing 
the vertices which contribute to  a given amplitude is also 
a graphical problem. In all these problems, the combinatoric element 
boils down to the counting of ribbon graphs, sometimes in large $N$ 
contexts and  sometimes not.

One motivation for studying matrix models (more specifically the 
Gaussian Hermitian and complex matrix models), which can be viewed 
as QFTs in {\it zero dimensions}, is that their correlators are related 
very closely to the combinatorics of ribbon graphs. This was reviewed 
in section 2.  It is well known, since 't Hooft's argument for the stringy 
structure of the large $N$ expansion, and the discovery of  low dimensional 
string duals for  various double-scaled limits of matrix models, that 
aside from the zero-dimensional structure, there is a { \it two-dimensional 
structure}  (related to string worldsheets) to this combinatorics. 
In the mathematical literature \cite{HarerZagier} the combinatorics 
of the Gaussian Matrix model shows up in the computation of the 
Euler character of moduli spaces of Riemann surfaces. In a recent 
revisiting of the Gaussian matrix model \cite{MR1}, 
the direct connection between 
Matrix model correlators (without double scaling limit) 
and an easily formulated question about counting triples of permutations, 
was used, along with the connection between such counting and 
branched covers of the sphere, to propose that there is a formulation of 
the Matrix model as a topological string theory with a $\mP^1$ target. 
Some explicit relations with the  the topological A-model
 string correlators were found in \cite{gopak}.

The initial goal of this paper was to formulate some counting 
problems related to ribbon graphs, motivated by QFT applications, 
and use permutation group techniques to solve these. 
We distinguished the counting with inverse automorphisms 
and the counting without inverse automorphisms, where the former
was directly related to correlators of the matrix models.

We used the mathematics literature which exploits Burnside Lemma and Hall's
theorem.
In section 3 this line of thought culminates in a formula for the number of
equivalence
classes of triples ${\cal N}(T_1,T_2,T_3)$ specified by the conjugacy
classes
$T_1$, $T_2$ and $T_3$. This is a refinement of the usual counting problem
in
that only ribbon graphs with a specified vertex and face structure are
counted.
Our treatment of this refined counting extends results that have appeared
in the
mathematics literature\cite{nedela}.
The counting in \cite{nedela} is achieved by reducing it to two distinct
problems: the problem of computing
the number of epimorphisms from the fundamental group of a Riemann surface
to $Z_l$
and the problem of computing the number of labelled graphs $\bar{L}$ in a
given equivalence
class $\big[\bar{L}\big]$, where $\bar{L}$ is obtained from $L$ by
quotienting with $Z_l$
(see the discussion following (\ref{useHall})).
To accomplish the refined counting we again need to count the number of
epimorphisms and
the number of labelled graphs in $\big[\bar{L}\big]$, but in addition we
have introduced a new
ingredient: the index distribution factor. 

In section 4, using the quotienting  method, we are able to give rather explicit
results for the refined
counting of several infinite families of ribbon and bi-partite graphs, with specified vertex and
face structure.
This section provides very concrete examples of how our methods work in
practice. The formulae in sections \ref{chords},\ref{section:6p}, after simplifications involving 
non-trivial binomial identities for character sums, match those found by different counting methods in the mathematics literature \cite{chord}\cite{triangulation}.  The other parts 
of section \ref{sequence} include new  infinite sequences.

In section 5 we further generalize our
counting results by counting Feynman graphs that have external legs. This
is a rather
interesting extension since the external legs are distinguishable. To deal
with this we
have introduced the notion of external-edge-labelled (EEL) graphs. EEL
graphs can be
put into one-to-one correspondence with a double coset. A straight forward
application of
Hall's theorem gives a remarkably simple formula for the number of EEL
graphs as a sum
over characters. We have also found it useful to introduce a second double
coset, obtained
by treating the external legs as indistinguishable.
This second coset is directly relevant to counting ribbon graphs that
appear in certain
correlators in the complex matrix model.
We develop an interesting relation between these two cosets.

One of the surprising results of this paper, 
is  a geometrical interpretation of the counting  in terms of { \it three dimensional } 
topological field theory. This can be viewed as counting 
maps from  membranes to a 3-manifold  $ \Sigma \times S^1$, 
where $ \Sigma $ is a 3-punctured sphere.

This work raises a number of questions and avenues for future research.

Some  generalizations of counting which should be accessible 
with the methods we have used involve triples of conjugacy classes  
\bea 
 [ \sigma  ]  = \{  [d]  ,[ 2^M 1^E ] ,[  n_1,n_2 , \cdots n_m ] \} 
\eea
where $d = 2M + E = n_1 + \cdots + n_m$, with  $m$ small. 
Another extension to consider is where  one of the 
conjugacy classes $T_i$ have just two cycles as opposed to one.  This 
will require the summation of characters  over irreps of $S_d$  corresponding to 
 Young diagrams with no more than $2$ boxes along the diagonal. 
This would be more elaborate but should be doable. 

The application of MHV rules to $\mathcal{N}=4$ SYM leads, through the use 
of a superspace approach, to the combinatorics of ordinary ribbon 
graphs.  MHV rules for less super-symmetric theories 
motivate the consideration of some other counting problems, 
where there are $2$ inward arrows at each vertex and some 
number of outward arrows. This should also be expressible 
in the language of permutation triples. Recent work on leading singularities 
of amplitudes have uncovered a connection to Grassmannians~\cite{ArkaniHamed:2009dn}, and 
graphs with two types of vertices, one three-point MHV and one three-point anti-MHV, so-called on-shell diagrams~\cite{nima}. 
These provide interesting extensions of the basic problem of 
counting bi-partite graphs, which can be encoded in the language of 
permutation triples and should be accessible with the methods of 
Burnside Lemma and graph quotients used here.

Our most tractable examples of ribbon graph counting involve a single face, which gives the 
lowest power of $N$, hence maximal genus contributions. 
In physics we have exploited the large $N$ expansion 
where the simplicity of the planar limit is used  to calculate the 
leading term. Is there a small $N$ expansion  ($N \rightarrow 0$)  where  the 
maximal genus contributions can form the leading terms of 
a systematic expansion? The simplicity of maximal genus has already been 
recognized in topological graph theory, in studies of the range of genera 
of the ribbon graphs which correspond to a given graph. 

We have refined ribbon graph counting - to keep track of more information 
beyond vertex structure. If we just keep track of vertex structure, closely analogous formulae exist for ordinary graphs (where the local symmetries of the 
vertices are symmetric groups as opposed to cyclic groups), which 
are relevant in QFT without large $N$ \cite{MR2}. Are there
 refined counting formulae  for these graphs analogous to the ones 
we have developed here? For example, one may refine according to degrees of divergence.

Using the connection between bi-partite graphs and 
classification of $N=1$ gauge theories mentioned in section 
\ref{sec:bipAds}, the reverse-quotient  construction  we have used here becomes 
 equivalent, in some special cases,
  to  the procedure   of orbifolding the space transverse to branes 
in the physics \cite{HK2005}. It is natural to ask if the more 
general quotient construction (and its reverse) of bi-partite graphs, 
 has an application in the context of quiver gauge theories or
quiver quantum mechanics models. The work of \cite{franco} which develops a construction of gauge theories  
for bi-partite graphs of any genus is a natural set-up for  exploring  this question.

We find the connection between the combinatoric element of 
QFT correlator/amplitudes  computations and 2D TFT as well as 3D TFT with 
$S_d$ gauge groups, with interpretations in terms of string worldsheet 
and membrane worldvolume maps particularly intriguing. It will be interesting 
to explore the  possible relation  of these hidden geometries,  
emerging from the combinatorics of  QFTs, to  geometries that
 arise  from physical duals (such as 
AdS/CFT or possibly some version of twistor strings) for the full QFT correlators/amplitudes.

\section*{Acknowledgements}

We thank  David Garner, Vishnu Jejjala, Hui Luo, Mingxing Luo, Jurgis Pasukonis, Gabriele Travaglini for stimulating discussions. 
CW would like to thank Peking University, Shanghai Jiaotong University and Zhejiang University for the hospitality where part of the work was done.
SR and CW are supported by  STFC Grant ST/J000469/1, String theory, gauge theory, and duality. 
RdMK is supported by the South African Research Chairs Initiative of the Department of Science and Technology  and National Research Foundation.

\begin{appendix}

\section{ Glossary } 

In this section we give a brief summary of the key terms and concepts used.

\begin{itemize} 

\item 
{\bf Permutation triples }\par
A collection of three permutations $\sigma_1,\,\sigma_2,\,\sigma_3\in S_d$ 
that satisfy $\sigma_1\sigma_2\sigma_3 =1$. Two triples  $(\sigma_1,\sigma_2,\sigma_3)$ 
and $(\sigma_1',\sigma_2',\sigma_3')$ are defined to be equivalent if 
\bea\label{equivtriples1} 
  \sigma_i'=\gamma\sigma_i\gamma^{-1}
\eea
for some $\gamma \in S_d$. 
These triples come up in the calculation of Matrix model correlators 
and in the description of bi-partite graphs  embedded on surfaces. 
See \ref{matbip} for explanation of  these connections.

\item 
{\bf Double-line diagrams }\par
Double line diagrams are used to describe the Feynman rules of quantum field theories (QFTs) 
with matrix fields, a special case being Matrix models which are QFTs in zero dimensions. 
The double lines  keep  track of  the row and column indices of the matrices. 
Each line can understood as the boundary of a strip on one side and a disc on the other.
Each Feynman diagram is then a collection of disks attached to each other by thin strips, to form
an oriented surface.  The weight of the Feynman diagram contains a
power of $N$ equal to the number of discs. In the large $N$ limit, these oriented surfaces were 
proposed to be worldsheets of an underlying string theory \cite{'tHooft}. This mechanism for the emergence of strings 
in large $N$ QFTs  underlies the AdS/CFT correspondence \cite{AdSCFT}.

\item
{\bf Ribbon graph}\par 
By shrinking the strips of the double-line diagram, we have a graph, consisting 
of vertices and edges, embedded on a Riemann surface (see Figure \ref{fig:dbrib}). 
The orientation of the Riemann surface induces a cyclic order at each vertex. 
Ribbon graphs can be defined in terms of edges and vertices, with a cyclic order at the vertices.

\item 
{\bf Bi-partite graph}\par
A graph, embedded on a surface, 
 whose vertices can be divided into two disjoint sets $V_b$ and $V_w$ such that
 every edge connects a vertex in 
$V_b$ to one in $V_w$.  We refer to vertices in these two sets as black and white vertices. 
Given any ribbon graph we can call its vertices black and introduce 
white vertices in the middle of each edge. This gives a bi-partite graph 
where all the white vertices are bivalent. 

\item
{\bf Permutation triples and bipartite graphs}\par 
 Choosing a labelling of the edges of a bi-partite graph with integers $\{ 1 , \cdots , d \}$ and 
 going around the black vertices according to the cyclic orientation, gives a collection of cycles, 
which form a permutation $\sigma_1$. Similarly going round the white vertices gives $\s_2$. 
The third permutation $\s_3 = (\s_1 \s_2)^{-1} $ gives information about the faces (i.e 2-cells 
in the complement of the graph). Its number of cycles is the number of faces. Different permutation 
triples related by conjugation (\ref{equivtriples1})  as above define the same bi-partite graph, which can be identified combinatorially as equivalence classes of these triples. A ribbon graph corresponds to bi-partite 
graphs where $\s_2$ is in the conjugacy class with $n$ cycles of length $2$, with $d=2n$. 
We write this as $[\s_2] = [2^n] $. The terminology {\it embedded graphs} includes both
ribbon graphs and bi-partite graphs.

\item 
{ \bf Labelled bi-partite graph} \par 
It can be identified with a permutation triple. We write $L = \{ \s_1 , \s_2 , \s_3 \} $. 
The bi-partite graph can be identified with the equivalence class of $L$ ( denoted $[L]$) under the equivalence (\ref{equivtriples1}).

\item 
{\bf Automorphism of an embedded graph} \par 
A permutation $\gamma \in S_d $ such that $\gamma \s_i \gamma^{-1} = \s_i$, for all
$i=1,2,3$. 
All the automorphisms of a triple form a subgroup of $S_d$  called the automorphism group 
($\Aut ( L ) $)
of the triple. Triples related by conjugation have conjugate automorphism groups. 
If  a permutation $\gamma$ is in the automorphim group of a triple $L = \{ \s_1 , \s_2 , \s_3\}$, 
this is also expressed by saying that  $\gamma$ is a stabilizer of $L$, or that $\gamma$ fixes $L$.

\item 
{\bf Burnside Lemma} \hfill\par
The Burnside lemma is a result from combinatorics, used to count orbits of a group action.
Let the finite group $G$ act on a set $X$. For each $g$ in $G$, $X^g$ denotes the set 
of elements in $X$ that are fixed by $g$. According to Burnside's lemma, the number of 
orbits is
\bea
   \hbox{ Number of orbits of the $G$-action on $X$}  = \frac{1}{|G|}\sum_{g \in G}|X^g|.
\eea
Thus the number of orbits  is equal to the average number of points fixed by an element of $G$. 
Useful references for the Burnside Lemma are  \cite{Cameron} \cite{wiki:burnside}.

\item 
{\bf Quotient of bi-partite graph }\par
Given a bi-partite graph $[L]$  with a representative  $L=\{ \s_1 , \s_2 , \s_3 \}$ and an automorphism $\gamma$ of $L$, there is a quotient $\bar L = \{ \bar \s_1 , \bar \s_2 , \bar \s_3 \}$. Hall's theorem ( Appendix \ref{sec:Hall}) implies that the conjugacy class of $\gamma$ in $S_d$ is $[\gamma ] = l^{ \bar d } $ for divisors $ l , \bar d $ of $d = l \bar d$. 
The quotient graph $[\bar L ] $ has $\bar d $ edges i.e $\bar \s_i \in S_{\bar d } $.

\item
{\bf Indices and epimorphisms related to a quotient} \par 
When we perform a quotient of the labelled graph  $L$ by a permutation $\gamma $ with $[ \gamma ] = l^{ \bar d } $
to get  the labelled graph $\bar L= \{  \bar \s_1 , \bar \s_2 , \bar \s_3  \} $, 
the cycles of $\gamma$  are denoted  $\{ S_1 \cdots S_{\bar d }\}$.  The relation between the cycles in  $\bar \s_i$  and those 
in $\s_i$ defines a set of integers $\nu_{i,j}$, one for each cycle in $\bar \s_i$ (see equation
(\ref{indexofcycle}) ). These indices enter an epimorphism equation (\ref{epieqn})
which is used in the counting of embedded graphs $[L]$. The number of solutions to this equation contributes an {\it epimorphism factor} to the main counting formula (\ref{genform}). 

\item 
{\bf Index distribution factor }\par
The index distribution factor appears when different cycles of the same length in 
a given permutation $\bar \s_i$ have the same index. The detailed form of this 
factor is explained in Section \ref{QuotsAndRev} and illustrated in examples 
in section \ref{App:exampsNCF}.

\item
{\bf Covering maps  and homomorphisms from fundamental group }\par
There is a correspondence between homomorphisms from $\pi_1(X) $, the fundamental group 
of a space $X$, to permutation groups $S_d$ and $d$-fold covers of $X$
(see standard texts in algebraic topology such as \cite{cst,hatcher} for the precise statements). This correspondence is central in the string theory of large $N$  2d Yang Mills 
theory and is reviewed in \cite{cmr}. It is used to relate permutation triples (hence 
matrix model correlators) to branched covers of a sphere, branched over precisely $3$ points in \cite{MR1}. It relates the counting of points on a double coset of permutation 
groups to covers of a cylinder or a torus, with appropriate restrictions on the 
monodromies \cite{MR2}.   We have used it here to relate $\cN(T_1, T_2 , T_3)$ 
to coverings a 3-manifold with boundary in section \ref{TFT}.  A variation on the correspondence 
replaces $S_d$  with a more general group. With a  $Z_l$ group, 
 it is  used in \cite{nedela1,nedela2,nedela} to arrive at  the epimorphism equation (\ref{epieqn}) as a key element in the counting of ribbon graphs via the quotient construction, as explained in section \ref{QuotsAndRev}.   

% Equivalence classes of homomorphisms from the fundamental group of a Riemann surface to %$Z_l$ are in
%1-1 correspodence with equivalence classes of branched covers.
% Recall that the homotopy group $\pi_1$ of a Riemann surface can be presented in terms of % % generators and relations 
 % as described in equation (\ref{epieqn}). Two homomorphisms $\pi_1\to Z_l$ are equivalent if % they differ
% by an inner automorphism of $Z_l$.
% The covers $f_1$ and $f_2$ are equivalent if there exists a homeomorphism such that % % $f_1\circ\phi =f_2$.
% The equivalence between homomorphisms and covers follows from a basic result in covering % space theory\cite{cst} 
% which establishes a 1-1 
% correspondence between conjugacy classes of subgroups of $\pi_1 (S)$ and equivalence % classes of topological 
% coverings of the space $S$.

\item 
{\bf Epimorphisms }\par
An epimorphism from a group $G$ to a group $H$ is a map $\phi:G\to H$ 
which is a group homomorphism and is surjective (onto). See \cite{epi}
for more details. Epimorphisms from the fundamental group 
of punctured Riemann surfaces to cyclic groups of order $l$, denoted 
$Z_l$, enter the counting described in section \ref{QuotsAndRev}.

\item 
{\bf Maps }\par
A map is synonymous with ribbon graph as described above. We will not use this terminology 
much, but it is common in a lot of the mathematics literature. A {\it labelled map}  (or labelled 
ribbon graph) with $n$ edges 
 is described as  a  labelled bi-partite  graph $ L= \{ \s_1 , \s_2 ,\s_3 \} $ with $\s_i \in S_d$ ; 
 $d= 2n $ , and conjugacy class of $\s_2 $ in $S_{d} $ given by $[\s_2] = [2^n]$. 
 A {\it rooted map } is endowed with  a distinguished edge called a root. The automorphism group of a (connected) rooted map is trivial.  Since the automorphism group of a labelled map is also trivial,  each rooted map with $d$ edges gives rise to $(d-1)!$ labelled maps. 
The number of rooted maps in an equivalence class of maps $[L]$  is the { \it number of orbits} 
of the automorphism group $\Aut ( L) $ acting on $\{ 1, \cdots , d \}$.  

\end{itemize}

\section{Some useful characters  } \label{charactersApen}

\subsection{ Character  for $[2^p] $ in hook representations }
Here we would like to calculate 
the characters of $[2^p]$ in the hook representation $R = [ 2p-k , k ] $.  
This can be done using the Murnaghan-Nakayama (MN)   Lemma (see \cite{fultonharris}).

There are two cases to consider. 
Case I is $ R = [ 2p- 2l , 1^{2l}] $ and Case II is 
$R = [ 2p-2l-1 , 1^{2l+1} ] $.  Let us start with Case I.
In applying MN lemma, we knock boxes $[2]$ or $[1,1]$
from $R$. We will knock off a total of $l$ copies of $[1,1]$ 
and $p- l $ copies of $[2]$. The last one we knock off has to be $[2]$ 
in this case. We sum over the possible sequences of knock-offs. 
Each sequence contributes a number which is the product of factors 
for each knock-off. The factor for  $[2]$ is $1$, and for
 $[1,1]$ it is $(-1)$. For any of these sequences we get 
\bea 
(-1)^l (1)^{ p -l } = (-1)^l  
\eea
The knock-off sequences can be labelled as 
\bea 
\{ a_1 , a_2 , \cdots , a_{p -1} , 1 \}
\eea
where $a_i = \pm 1 $. The weight of the sequence 
is $\prod_{i} a_i = (-1)^l $. The number of these sequences is 
$  { p-1 \choose l  }  $. So we conclude that 
\bea 
\chi_{ [2p -2l , 1^{2l } ]} ([2^p ]  ) = (-1)^l  { p-1 \choose l  } 
\eea

For Case II, the last hook we knock off has to be $[1,1]$. So in this case 
the sequences look like 
\bea 
\{ a_1 , a_2 ,\cdots , a_{p - 1} , -1 \} 
\eea
and the weight is always $(-1)^{l+1} $. There are $l$ copies of $(-1)$ 
among the $a_1 \cdots a_{p -1}$. 
We conclude that
\bea 
\chi_{ [2p-2l-1, 1^{2l+1} ] }  ( [2^n] ) =  (-1)^{l+1}  
{ p -1 \choose l   }  \, .
\eea
Alternatively both results may be expressed in an unified form
\bea\label{chi2topower} 
\chi_{ [2p- k , 1^{k } ] }  ( [2^p] ) =  (-1)^{ \lfloor (k +1)/2 \rfloor }  
{ p-1 \choose \lfloor  k/2  \rfloor  } \, .
\eea
where $\lfloor x \rfloor$ means we take the integer part of $x$.
The following sum of characters normalized by dimensions, taken over hook representations $[2p-k , 1^k  ]$ has a simple form 
\bea
\sum_{R}  { \chi_{R}([2^p])  \over d_{R} } &=& 0 ~~ \hbox{ for $p$ odd } \cr 
& = &  {  2p  \over p +1 }   ~~ \hbox{ for $p$ even }  
\eea
This is a special case of  (\ref{IDpq}).

\subsection{ Character for $ [ 2^p , 1^{q} ] $  in hook representation } 

We will assume $ q \ge 1 $, and let $n=2p+q$. 
Suppose, in applying the MN lemma, we knock off 
$[2^l ] $ from the first row of $R$, and $ [ 2^{p-l } ] $ 
from the column. The weight is $(-1)^{ p - l } $ and a combinatoric factor 
of 
\bea 
 { p \choose l } 
\eea
We have to sum over different possible values of $l$. 
The upper  limit is obtained as 
\bea 
2 l  \sim   ( n - k -1  ) 
\eea
If $ n -k-1 $ is even the upper limit is precisely 
\bea 
l  = { ( n - k - 1 ) \over 2 } 
\eea
If $ ( n -k -1)$ is odd, then upper limit is 
\bea 
l = { ( n- k -2 ) \over 2 } 
\eea
Both cases can be expressed as 
\bea 
l =  \lfloor  { ( n - k - 1 ) \over 2 } \rfloor 
\eea
Consider the lower limit. Now we need to consider 
\bea 
( 2p - 2l )  \sim k 
\eea
If $k$ is even, we have $ l = p - k/2$. If $k$  is odd, 
the lower limit is at $ 2p - 2l = k -1 $, i.e $ l = p - k/2 +1/2 $.
So lower limit is 
\bea 
l = \lfloor p - k/2 +1/2  \rfloor 
\eea 

Once we have knocked off the $2$'s, we can knock off the 
$1$'s. The last $1$ can be removed in only one way - this is where the assumption 
$q>0$ plays an important role. 
The remaining $q-1$ copies of 1 can be removed from either 
the column or the row, which can be described by a 
sequence of length $q-1$ of $c$ and $r$, denoting the choice of 
column or row. So we get 
\bea 
{  q-1 \choose   k - 2p + 2l  } 
\eea
Hence we can write   
\bea\label{char2p1q} 
\chi_{[n-k , 1^k ]  } ( [2^p,1^{q} ]  ) 
&=& \sum_{ l = \lfloor p - k/2 +1/2   \rfloor   }^{ \lfloor ( n - k -1 )/2 \rfloor    }  
(-1)^{ p - l  }  { p \choose l }  
{  q-1 \choose   k - 2p + 2l  } \cr 
&=&  \sum_{ l = 0    }^{ p  }  
(-1)^{ p - l  }  { p \choose l  }  
{  q-1 \choose   k - 2p + 2l  }
\eea
In the last line, we have simplified the summation range by adding terms that vanish 
due to the zeroes of $ { p \choose l  }  = {  \Gamma ( p+1 ) \over \Gamma ( l+1 ) \Gamma ( p -l +1 ) }$. 
In the special case $q=0$, we can use the formula previously derived (\ref{chi2topower}).
An expression covering all cases can be written as follows 
\bea\label{form1}  
 && \chi_{[n-k , 1^k ]  } ( [2^p,1^{q} ]  ) \cr 
    &&                    = \left [   ( 1 - \delta_{ q , 0 } )   \sum_{ l = 0    }^{  p     }  
(-1)^{ p - l  }  { p \choose l }   {  q-1 \choose   k - 2p + 2l  }   
 + \delta_{ q,0}   (-1)^{ \lfloor (k +1)/2 \rfloor }  
{ 2p -1 \choose \lfloor  k/2  \rfloor  }   \right ]   
\eea

The same character, $\chi_{ [n-k , 1^k] } ( [2^p , 1^{q}] ) $ can be computed 
 by applying MN lemma, with a choice of  knocking off $1$'s first, and then follow it with the  $2$'s. 
If we knock off $l$ of the $1$'s from the first column, there is a combinatoric factor 
of  ${ q \choose l }$ ways of choosing  these $1$'s. After this is done we are left with $[2^p]$ in representation 
$[2p-k+l, 1^{k-l}]$, which leads to 
\bea
(-1)^{ \lfloor (k-l+1)/2 \rfloor } {  p-1 \choose \lfloor  (k-l)/2 \rfloor   } \, .
\eea
Putting these  together, we have 
\bea
\chi_{[n-k, 1^k]}([2^p, 1^q]) = \sum^{k}_{ l =0 } (-1)^{ \lfloor (k-l+1)/2 \rfloor } { q \choose l }  {  p-1 \choose \lfloor  (k-l)/2 \rfloor   }
\eea 
Here we have assumed that $ p > 0 $. For the special case $p=0$, the character is given as
\bea \label{character[1n]}
\chi_{[n-k, 1^k]}([ 1^n ] ) = { n - 1 \choose k } \, . 
\eea
From this approach we arrive at 
\bea\label{form2}  
&& \chi_{[n-k , 1^k ]  } ( [2^p,1^{q} ]  ) \cr 
 &&   =  \left [  ( 1 - \delta_{ p  , 0 } )  \sum^{k}_{ l =0 } (-1)^{ \lfloor (k-l+1)/2 \rfloor } { q \choose l }  {  p-1 \choose \lfloor  (k-l)/2 \rfloor   } 
  ~~~ + \delta_{ p,0}  { n - 1 \choose k } \right ]   
\eea

Using either of the equivalent expressions, (\ref{form1}) or (\ref{form2}),  one can check, by inserting explicit 
values of $k,p,q$ (e.g in Mathematica) with $n \equiv 2p + q $, that the sum over hook representations $R = [ n-k , 1^k  ] $ gives 
 \bea\label{IDpq} 
 \sum_R { \chi_R ( [2^p, 1^q] ) \over d_R }  & = &  {  ( 2p +q )  \over (p+1) }    ~~~ \hbox{ for $p$ even }  \cr 
                                                                        & = & 0  ~~~ \hbox{ for $p$ odd }  
 \eea
The $p$ odd case is easy to understand using the fact that the characters of a permutation 
in  conjugate representations are equal up to a sign given by the parity of the permutation, so the sum 
over all hooks, which is self-conjugate, vanishes. The $p$ even case is more non-trivial. 
 We do not have a direct analytical proof of this, but the simple form was guessed by comparing
our character based approach for counting bi-partite graphs with $ [ \bar \sigma ] = \{ [4p ] , [2^{2p} ] , [4p]  \}$
 with  the equivalent chord diagram counting ( section \ref{chords} ).

\subsection{Character for $ [ 3^p , 1^q ] $  in hook representation }
Similarly we can obtain the character for $ [ 3^p , 1^q ] $  in $ R= [ n-k , 1^k] $
\bea   \label{3p1q}
\chi_{R}([ 3^p , 1^q ] ) 
=
\left [  ( 1 - \delta_{ p  , 0 } )  \sum^{ k}_{t=0 }  (-1)^{{\lfloor k/3 \rfloor} + k} {q \choose t} 
{p -  1 \choose \lfloor(k - t)/3\rfloor  } 
  ~~~ + \delta_{ p,0}  { n - 1 \choose k } \right ]
\eea
where $n = 3p + q$. The formula can be derived by knocking off $q$ $1$'s first, and following it by knocking off $p$ $3$'s in the representation $R = [n-k, 1^k]$. Combinatoric factor of knocking off $t$ $1$'s from row and $(q-t)$ $1$'s from column is given as
\bea
{ q \choose t} \, .
\eea
After remove all the $1$'s, we then knock off $p$ $3$'s, it gives
\bea
(-1)^{{\lfloor k/3 \rfloor} + k} { p-1 \choose \lfloor (k-t)/3 \rfloor  } \, ,
\eea
where the sign factor $(-1)^{{\lfloor k/3 \rfloor} + k}$ is obtained by examining different $k$ module $3$. Put them together and sum over $t$ we obtain the character (\ref{3p1q}). Here $ p > 0 $ is assumed in the derivation, for the special case $p=0$, the character is given in equation (\ref{character[1n]}), as we summarized in Eq.~(\ref{3p1q}). Interestingly, as we noted in section \ref{section:6p}, this character satisfies various highly non-trivial identities, see equations (\ref{6p-Z1}), (\ref{6p-Z2}), (\ref{simp2}) and (\ref{simp3}).

\section{ On calculation of the number of epimorphisms } \label{appendix:N_n}

Here we give a simple derivation for the number of solutions of the epimorphism equation which appears in section \ref{sequence}
\bea 
&& e_1^3 = e_2^3 = \cdots = e_n^3 = 1 \cr 
&& e_1 e_2 \cdots e_n =1 
\eea
where $e_i \in Z_3$. We will use methods of the Fourier transform 
on finite groups (see \cite{joneshoms} for a general discussion)

Let $ \omega = e^{ 2 \pi i \over 3 }$. 
Each $e_i $ is $\omega^{a_i}$ where $a_i \in \{ 1 , 2 \} $. 
Recall that 
\bea 
\delta ( e_1 \cdots e_n )   = \sum_{ R \in Reps ( Z_3 ) } { d_R \chi_R ( \prod e_i  ) \over 3 } 
\eea
There are 3 irreps of $Z_3$, all one dimensional -- call them $R_0, R_1 , R_2$. 
Let $g$ be the generator of $Z_3$ obeying $g^3 =1 $. The action of $g$ 
in the three      irreps is given by 
\bea 
&& g |R_0 \rangle = |R_0 \rangle \cr 
&& g |R_1 \rangle = \omega |R_1 \rangle \cr 
&& g |R_2 \rangle = \omega^{-1}  |R_2  \rangle  
\eea
So we have 
\bea 
\delta ( e_1 \cdots e_n )  = { 1 \over 3 } + { 1 \over 3 } \omega^{ a_1 + \cdots + a_n } + { 1 \over 3 }  \omega^{ - a_1  - \cdots  -  a_n }
\eea
Suppose $k$ of the $a$'s are $1$ and the rest two, 
then 
\bea 
\delta ( e_1 \cdots e_n )  = { 1 \over 3 } + { 1 \over 3 } \omega^{
 2k - n  } + { 1 \over 3 }  \omega^{ n - 2k  }
\eea 
There are $ { n \choose k } $ possibilities of which $a$'s are
equal to $1$. So we get the number of solutions for the epimorphism equation
\begin{equation} 
\begin{split} 
N_n & = \sum_{ k=0}^n { n \choose k }  { 1 \over 3 }  ( 1 +   \omega^{
 2k - n  } +  \omega^{ n - 2k  } ) \\
& = { 2^n \over 3 } + { 2\over 3 }  Re  ( \omega^{ n } ( 1 + \omega^{-2} )^n )
 \\ 
& =   { 2^n \over 3 } +  { 2\over 3 } Re ( \omega + \omega^{-1}  )^n  \\ 
& =  { 2^n \over 3 } + { 2 \over 3 } Re ( 2 Cos ( { 2 \pi \over 3}  ) ) ^n  \\
& =   { 2^n \over 3 }  +  { 2 \over 3 } (-1)^n 
\end{split}
\end{equation}

\section{   Hall's theorem    }\label{sec:Hall}  

Hall's  theorem is a very useful fact from the theory of embedded 
graphs which constrains the automorphisms of these graphs, 
or equivalently of permutation triples \cite{nedela,Liskovets1,Liskovets2}. 
 Let G be the group generated by $\s_1 , \s_2$ ;
 For any pair $i , j$  in  the set of edges $\{ 1, \cdots ,  d \}$ 
 there is a $\s \in G$  such that
$ \s ( i ) = j$
i.e.  $\s$ acting on $i $ gives $j$ (this is the  transitivity condition which is equivalent
to connectedness).

Let $C$ be the subgoup of $S_d$ which commutes with $G$.
Let $\gamma$ be an element of $C$  ( i.e something in the automorphism group of the ribbon graph).
If an integer $i$ belongs to some cycle of length $l$ in $\gamma$, this means
 \bea\label{gkii} 
\gamma^l ( i ) = i 
\eea
 and no other smaller power of $\gamma$ leaves $i$ fixed.

 Using  \ref{gkii}  let us show that
\bea 
   \gamma^l (j ) = j
\eea
which would establish that $j$ belongs to a cycle of the same length $l$.

From transitivity we know there is a $\s$ such that
\bea 
 j = \s (i )
\eea
So (\ref{gkii}) implies 
\bea 
 \gamma^l  \s^{-1} ( j ) = \s^{-1} ( j ) 
\eea
Hence
\bea  
\s \gamma^l  \s^{-1}  (j ) = j
\eea
But $\s$ and $\gamma$ commute, so we learn 
\bea 
   \gamma^l  ( j ) = j
\eea

And this must be the smallest power of $\gamma$ which fixes $j$. 
 If there was some smaller power $l' < l$ of $\gamma$  that obeyed
\bea 
   \gamma^{l' } ( j ) = j
\eea
then by running the above argument backwards we would learn
 that
\bea 
   \gamma^{ l' } (i) = i
\eea
 which would contradict the assumption that $l$ was the smallest power of
 $\gamma$ that fixes $i$.

So we have Hall's theorem.  All the  integers between $ \{ 1 , \cdots , d \}$
 belong to cycles of the same length $l$ in any element $\gamma$ of the
Automorphism group. Equivalently the cycle structure of $\gamma$ is 
of the form $[l^{\bar d} ]$  for a pair of divisors  $(l , \bar d )$ obeying $l \bar d = d $. 

\section{ Quotienting  bi-partite graphs as an operation
 on permutation triples   } 
 \label{Appsec:quotienting}

We explained in section \ref{QuotsAndRev} that given a 
labelled graph $L = \{ \s_1 , \s_2 , \s_3\}$, and an automorphism 
$\gamma$,  we can obtain a quotient graph $\bar L$, indices 
which are integers associated with each cycle of the permutations 
$ \bar \s_i$, as well as epimorphisms given by group 
elements associated with the non-trivial cycles of the Riemann surface 
supporting $ \bar L$ as well as the cycles of $ \bar \s_i$. 
Let us explain how to read off the latter group elements.
 As an example suppose 
$\bar \s_1   $ contains a cycle of the form $( S_1 S_2 ) $. 
And suppose the index for this cycle is three. 
This means that there is a cycle in $\s_1$
of the form
 $(S_{1 , b_1  } S_{2 , c_1  }   
S_{1 , b_2  }  S_{2 , c_2  }  S_{1 , b_3  }  S_{2 , c_3 } )  $, such that 
 $ ( S_{1 , 1  } S_{1 , 2  }  S_{1 , 3  } ) $  and 
$ (  S_{2 , 1 }  S_{2 , 2  } S_{2 , 3 }    )     $ 
are cycles in $\gamma $ ; $( b_1 , b_2 , b_3 ) $ are a permutation 
of $(1,2,3)$ and likewise $( c_1 , c_2 , c_3 ) $ are a permutation 
of $ (1,2,3)$. The permutations obtained from the 
$b$'s is the same as the permutation from the $c$'s. So this is a 
 a plausible way to extract the exponents $m_j$ which specify the 
 epimorphism to $Z_l$. To see this prescription at work in a concrete 
example, see the equation (\ref{nuandep}).

\subsection{ Reverse of Quotient construction 
 and Angle Voltages   }\label{App:reverseconstruction} 
We have already described how to quotient a graph by a group of automorphisms $Z_l$ to obtain a reduced graph. 
We now want to consider the inverse of this process, that is, given a reduced graph and an automorphism group,
how do we recover the graph that gives the reduced graph after we quotient with the automorphism group? 
The graph we recover from the reduced graph and specified automorphisms is called the derived graph.
The key idea that we use is that of voltage assignments, as developed in \cite{voltage}.

Given a graph specified by permutations $\bar{\sigma}_1$ and $\bar{\sigma}_2$, an angle 
is defined by any pair of coincident edges i.e. 
$(S_x, S_y)$ with $S_y\in\{\bar{\sigma}_1 (S_x),\bar{\sigma}_1^{-1} (S_x),\bar{\sigma}_2 (S_x),\bar{\sigma}_2^{-1}(S_x)\}$. 
A voltage assignment $\alpha$ assigns an element of the group of automorphisms $Z_l$ to each angle at the vertices.

Denote the number of edges of the reduced graph by $\bar{d}$. The order of the automorphism group is $l$.
The derived graph has $\bar{d}l$ edges. Each edge can be labelled by a pair $(x,g)$ where $1\le x\le \bar{d}$ 
and $g\in Z_l$.
The action of the permutations specifying the derived graph are
\bea
   \sigma_i \left(S_{x,g}\right)=S_{\bar{\sigma}_i(x),g\cdot\alpha\left(x , \bar{\sigma}_i(x)\right)}
\eea

\begin{figure}[h]
\scalebox{1.0}{
\centerline{\includegraphics[height=4.5cm]{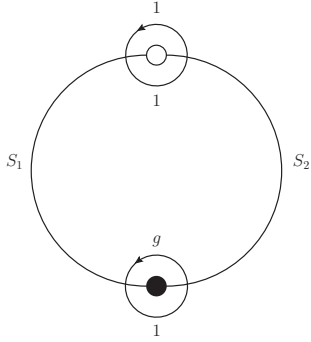} }
%\centerline{\includegraphics{triangle} } 
}
\caption{\it
Example of a voltage assignment. }
 \label{fig:fdiagram1}
 \end{figure}

An example to illustrate these rules is in order. Consider the reduced graph described by the permutations
\bea
  \bar{\sigma}_1 = (S_1 S_2)\qquad \bar{\sigma}_2 = (S_1 S_2)
\eea
We consider the automorphism group $Z_2$ with elements $\{1,g\}$. The voltage assignment we use is given in Figure \ref{fig:fdiagram1}.

%
%
%%%%%%%%%%%%%%%%%%
%%%%%%%%%%%%%%%%%%%
%
%

\begin{figure}[h]
\scalebox{1.0}{
\centerline{\includegraphics[height=4.5cm]{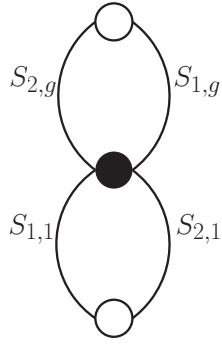} }
%\centerline{\includegraphics{triangle} } 
}
\caption{\it
Derived graph for the voltage assignment of Figure \ref{fig:fdiagram1}. }
 \label{derived}
 \end{figure}

The edges of the derived graph are $\{S_{1,1},S_{1,g},S_{2,1},S_{2,g}\}$.
Using the above voltage assignment we have
\bea
  \sigma_1 \left( S_{1,1}\right)=S_{\bar{\sigma}_1(1),1\cdot g}= S_{2,g}\cr
  \sigma_1 \left( S_{2,g}\right)=S_{\bar{\sigma}_1(2),g\cdot 1}= S_{1,g}\cr
  \sigma_1 \left( S_{1,g}\right)=S_{\bar{\sigma}_1(1),g\cdot g}= S_{2,1}\cr
  \sigma_1 \left( S_{2,1}\right)=S_{\bar{\sigma}_1(2),1\cdot 1}= S_{1,1}
\eea
and
\bea
  \sigma_2 \left( S_{1,1}\right)=S_{\bar{\sigma}_2(1),1\cdot 1}= S_{2,1}\cr
  \sigma_2 \left( S_{2,1}\right)=S_{\bar{\sigma}_2(2),1\cdot 1}= S_{1,1}\cr
  \sigma_2 \left( S_{1,g}\right)=S_{\bar{\sigma}_2(1),g\cdot 1}= S_{2,g}\cr
  \sigma_2 \left( S_{2,g}\right)=S_{\bar{\sigma}_2(2),g\cdot 1}= S_{1,g}
\eea
Thus, the derived graph has
\bea
  \sigma_1 = (S_{2,g} S_{1,g} S_{2,1} S_{1,1}),\quad
  \sigma_2=  (S_{1,1} S_{2,1})(S_{1,g}S_{2,g})
\eea
This graph is shown in Figure \ref{derived}.

%
%
%%%%%%%%%%%%%%%%%%
%%%%%%%%%%%%%%%%%%%
%
%

If we consider a different voltage assignment, assigning $1$ to every angle,
the derived graph is a disconnected graph given by two copies of the original, i.e.
the graph with
\bea
  \sigma_1 = (S_{1,1}S_{2,1})(S_{1,g}S_{2,g})\qquad \sigma_2 = (S_{1,1}S_{2,1})(S_{1,g}S_{2,g})
\eea

For the counting application we consider, it is clear that the derived graph must
be connected. It is therefore useful to understand the necessary condition on the voltage
assignment that gaurantees a connected derived graph. A useful way to prove that the
derived graph is connected is to prove that it is possible to pass from a given dart $S_{i,g_1}$
to any other dart $S_{j,g_2}$ without leaving the graph.

Since the reduced graph is connected, it is clear that there is a path on the graph from 
a given dart $S_{i,g_1}$ to another dart $S_{j,g_2}$ where we can choose $1\le i,j\le \bar{d}$ arbitrarily.
The voltage for any closed path on the graph
picks up a factor for each angle we pass through. By accumulating these factors we can assign a 
voltage $g_C$ to any closed path $C$, such that if we start from dart $S_{i,g}$ we pass to dart
$S_{i,g_{_C} \cdot g}$ when we traverse the path. Denote the group generated by the collection of $g_C$,
one for each path, by ${\cal A}_C$. To pass from $S_{i,g_1}$ to $S_{i,g_2}$ for any $g_1,g_2$
by traversing closed paths $C$, ${\cal A}_C$ must act transitively on $Z_l$ implying that
${\cal A}_C$ must be isomorphic to $Z_l$ itself.

\subsection{ Index distribution factor  } \label{App:exampsNCF}

We give examples of quotients  where the index distribution factor of 
(\ref{genform}) is non-trivial. 

\subsubsection{  A $Z_2$ cover of $\bar d = 2$ graph } 

\begin{figure}[h]
\scalebox{1.0}{
\centerline{\includegraphics[height=4.5cm]{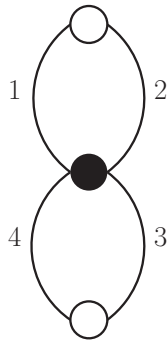} }
%\centerline{\includegraphics{triangle} } 
}
\caption{\it
Figure of Eight. }
 \label{fig:figOf8}
 \end{figure}

Consider the graph shown in Figure \ref{fig:figOf8}

%
%
%%%%%%%%%%%%%%%%%%
%%%%%%%%%%%%%%%%%%%
%
%

It is described by a permutation triple 
\bea 
 \s_1 = ( 1,2,3,4)  ~ ; ~ \s_2 = (1,2) ( 3,4 ) ~ ; ~ \s_3 = (1) (2,4) (3) 
\eea
The conjugacy classes are 
$ [ \sigma ] = \{   [4] , [2,2] , [2,1,1]  \} $. 

The automorphism group  $ \Aut ( L ) $  is  generated by $ ( 1,3 ) ( 2,4) $
and is isomorphic to $Z_2$. Labelling
these cycles  $ S_1 = (1,3) , S_2 = (2,4)$ and 
quotienting by this generator gives 
\bea 
&& \bar L : \bar \s_1 = ( S_1 S_2 ) ; \bar \s_2 = ( S_1 S_2 ) ; \bar \s_3 = 
(S_1) ( S_2) \cr  
&& \nu ( \bar L ) = \{ [ 2  ; 1  ; 1 , 2 ] \cr 
&& { \cal E } ( \bar L )  =  \{ [ g ; 1 ; 1 ; g  ] \} 
\eea
$\bar L$ is the labelled quotient graph, described in terms of 
a permutation triple. 
$\nu ( \bar L )$ gives indices. ${ \cal E } ( \bar L )$ gives
the actual group elements $e_i$, so $e_i^{\nu_i} =1 $. 
$g$ is the generator of $Z_2$.

There  are $12$ labelled graphs $L$, generated by conjugating 
the triple $ ( \s_1 , \s_2 , \s_3 )$ with
permutations in $S_{4}$. Of these, two are fixed  by a given $\gamma$. 
So the contribution to the Burnside sum 
for counting the number of unlabelled graphs 
with the specified $[\s ] $ is $1/2$. Another 
$1/2$ comes from the case where $\gamma $ is the identity permutation.

Note that, from the simple formula (\ref{simpform}), we get  
\bea 
{ 1 \over 4 } \times \hbox{ labelled quotient graphs  }  \times \hbox{ epimorphisms }    = 
{ 1 \over 4 }
\eea
which does not match the correct $1/2$. 
This is because there are $2$ unlike 
indices for 2 cycles of same length. 

The group $ G ( \nu ( \bar L ) )$ in this case is 
an $S_2$ of permutations of   the cycles $(1) , (3)$  
in $\bar \s_3$. These two cycles are also permuted 
by the element $(1,2)$ of $\Aut ( \bar L ) $.  So 
$ |\Aut ( \bar L ) \cap  G ( \nu ( \bar L ) )| = 2 $. 

So when the $S_m $ subgroup of $S_m [ Z_l ] $ 
acts on the labelled $L$'s fixed by a 
given $\gamma$, it acts on $ [ \bar L , \nu ( \bar L ) ] $. 
If we forget about the indices, we get all the 
labelled $\bar L $. But a given $\bar L$ can appear multiple times 
in the $S_m$ orbit on the pairs  $ [ \bar L , \nu ( \bar L ) ] $, 
when that happens it means that an element of $S_m$ which 
fixes $ \bar  L$ ( hence is in $\Aut ( \bar L ) $ ) performs 
a permutation of the like-cycles carrying unlike indices.

\subsubsection{ A   $Z_3$ cover  of a $\bar d =4 $ graph    } 

Take 
\bea 
&& \sigma_1 = ( 1,2)( 5,6) ( 3,11,7) ( 9,10 ) ( 4) (8) (12) \cr 
&& \sigma_2 = ( 1,4 ) ( 5,8 ) ( 9,12) ( 2,7,6,11,10, 3 ) \cr 
&& \sigma_3 = ( 1,4,2,7) ( 3,9,12,10) ( 5,8,6,11) \cr 
&& \Aut ( L )  = \{ (), 
( 1,5,9) ( 2,6,10 ) ( 3,7,11) ( 4,8,12) , 
( 1,9,5) ( 2,6,10 ) ( 3,11,7) ( 4,8,12) \} \cr 
&& ~~ 
\eea

Quotient by 
\bea 
\gamma_1 = ( 1,5,9) ( 2,6,10 ) ( 3,7,11) ( 4,8,12)
\eea 
The quotient graph is given by the triple 
\bea 
&& \bar{\s}_1 = ( S_1,S_2) (S_3) ( S_4) \cr  
&& \bar{\s}_2 = ( S_1,S_4)  ( S_2,S_3)  \cr 
&& \bar \s_3 = ( S_1,S_4,S_2,S_3)  
\eea
The automorphism group of the quotient graph is 
\bea 
\{ () , ( S_1,S_2) (S_3,S_4) \} 
\eea
The number of orbits of this on the $\{ S_1, S_2, S_3, S_4 \} $
is the number of rooted graphs. The index of this graph quotient is 
\bea\label{nuandep}  
&& \nu ( \bar L ) =  [ 1,3,1 ; 1,3 ; 1 ] \cr 
&& \cE (  \bar L ) = [ 1,g^{-1} ,1 ; 1,g ; 1 ] 
\eea
 As a side remark, note that we can read off the epimorphism 
by comparing the order of numbers in a cycle of $\gamma$ of non-trivial index
and the order in which these numbers appear in the lift of that cycle 
to $L$.

Let $ Symm ( \g_1)$ be the subgroup of $S_d$ which commutes with $\g_1$.  
Some general facts about $Symm(\g_1)$ orbits turn out to be very useful 
here.  The set of all labelled graphs fixed by $\gamma_1$ forms 
orbits of $Symm ( \gamma_1) = S_4[Z_3] $. This is because 
\bea 
 \gamma_1 L \gamma_1^{-1} && = L \cr 
 \mu \gamma_1 \mu^{-1} && = \gamma_1 \cr  
 \implies  \mu \gamma_1 \mu^{-1} L \mu \gamma_1^{-1}  \mu^{-1} && = L \cr  
 \gamma_1 ( \mu^{-1} L \mu )  \gamma_1^{-1} && = ( \mu^{-1} L \mu )  
\eea
All labelled graphs corresponding  to the same unlabelled graph 
are generated by $S_d = S_{12} $ action. Suppose we have 
$L_1, L_2 \cdots L_k$ are labelled graphs in the same $S_d$ orbit, 
all stablized by $\g_1$  but in distinct $Symm (\g_1) $ orbits. 
So we have, for any $j \in \{ 1 , \cdots , k \}  ; j \ne 1 $ 
\bea 
L_j = \alpha_{j} L_1 \alpha_{j}^{-1} \hbox{ for some } \alpha_{j }  \in S_d 
\eea
with $ \alpha_j \not\in Symm (\g_1)$.
We know that 
\bea 
&& \gamma_1 L_j \gamma_1^{-1} = L_j \cr 
&&  \gamma_1 \alpha_j^{-1}  L_1 \alpha_j \gamma_1^{-1} = L_1 \cr 
&&  \a_j  \gamma_1 \alpha_j^{-1}   L_1 \alpha_j \gamma_1^{-1} \a_j^{-1}
 =  \a_j L_1 \a_j^{-1}  
\eea
So we learn that $  \a_j  \gamma_1 \alpha_j^{-1} $ stabilizes $L_1$. 
Since $\a_j \not \in Symm ( \g_1) $, we can conclude that 
$ \gamma_j \equiv  \a_j  \gamma_1 \alpha_j^{-1} $ is a distinct element 
in $\Aut (L_1) $ which has the same cycle structure as $\gamma_1$.
For any pair $i,j \in \{ 1 , .. , k \} ~ ; ~ i , j \ne 1 $ we have 
\bea 
L_i = \alpha_{ij} L_j \a_{ij}^{-1 } 
\eea
 where $ \alpha_{ij} = \alpha_i \a_j^{-1} $. Since $L_i, L_j$ are in distinct 
$Symm(\g_1)$ orbits, we know that $\alpha_{ij} \not\in Symm (\g_1) $. 
This allows us to prove that $\gamma_i \ne \gamma_j $. Indeed 
if $\gamma_i = \gamma_j $, then 
\bea 
&& \alpha_i \gamma_1 \alpha_i^{-1} = \alpha_j \gamma_1 \alpha_j^{-1} \cr 
&& \implies \gamma_1 = \alpha_j^{-1} \alpha_i ( \gamma_1 ) \alpha_i^{-1}
 \alpha_j  
\eea
which contradicts the assumption  $\alpha_{ij} \not\in Symm (\g_1) $. 

We conclude that the number of distinct automorphisms 
of $L_1$ in the same conjugacy class as $\g_1$ is 
greater or equal to the number of orbits of $Symm (\g_1) $ among the 
labelled graphs fixed by $\g_1$.

We can apply this in the problem at hand. 
Using $Symm(\gamma_1)$ we generate a list of $648$ labelled graphs. 
We observe that $\gamma_2$ is obtained by conjugating $\gamma_1$ 
with $(5,9) ( 6,10) (11,7)( 8,12)$. We conjugate $L$ with this.
Then generate labelled graphs by $Symm(\g_1)$. We get another 
$648$. We look at the intersection of the two lists to find it 
is empty. This is conveniently done using GAP \cite{GAP}.  
 So we have found two orbits of $Symm(\g_1)$ 
and there can be no more since there are exactly two $\gamma$'s in 
this conjugacy class.

From Eq.(\ref{useHall}) the 
contribution to the Burnside sum  from this conjugacy  class of $\gamma$'s is 
\bea 
{ 1 \over l^{ \bar d }   { \bar d }!  } 648 \times 2 = 2/3 
\eea
There are  $12 $ labelled quotient graphs,  2 epimorphisms,
and 
\bea 
 \Aut ( \bar L ) \cap G (  \nu ( \bar L ) ) 
 = \Aut ( \bar L ) \cap ( S_2 \times S_2 )  = S_2 
\eea
So we calculate the contribution to Burnside sum using Eq.(\ref{genform})  is 
\bea 
{ 1 \over 3 \times 4!  }\times 2 \times 12  \times 
|\Aut ( \bar L ) \cap G (  \nu ( \bar L ) )  | = 2/3 
\eea
So this provides another non-trivial check of (\ref{genform}). 

\section{ Chord Diagram counting and ribbon graphs } \label{appendix:chords}

In section \ref{chords}  we made contact between our 
counting of ribbon graphs and the counting of chord diagrams. 
This connection can be nicely understood using the 
origin of the ribbon graphs in the Matrix model correlators
described in Section \ref{matbip}. 
Let us illustrate this with the example of the 1-point function of 
$ \langle tr ( \Phi^4 )   \rangle $ that we described there. 
We showed how to associate ribbon graphs with  the different Wick 
contractions by first going to double line diagrams and then 
thinning these down to get the ribbon graph. Now for every vertex of 
such a ribbon graph, we can surround it by a small circle, erase the 
vertex in the middle of the circle along with the edges connecting it to the 
circumference of the circle. Suppose the $i$'th vertex of the ribbon 
graph has $e_i$ edges connected to it. Applying the above procedure 
to it, results in an increase of the number of vertices by $e_i -1 $, 
since the vertex in the middle was removed, but vertices on the circumference 
were added. The number of edges increases by $e_i$. The number of faces 
increases by $1$. So the Euler character does not change. 
If the ribbon graph is described by permutation triple 
$( \s_1 , \s_2 , \s_3 )$, where $C_{ \s_1} $ is the number of vertices, 
$C_{ \s_2} $ the number of edges, and $C_{ \s_3} $ the number of faces, in the new graph
 we have $C_{ \s_1} + C_{ \s_2}  - C_{ \s_1}  = C_{ \s_2} $ vertices, 
$2C_{ \s_2}   $  edges and 
$C_{ \s_1}  + C_{ \s_3} $ faces.

For 1-vertex ribbon graphs, this transformation produces chord diagrams. 
For multi-vertex ribbon graphs, we still have points on circles connected 
by edges, embedded on the Riemann surface. This transformed diagram  
can be understood in the Matrix theory calculation as a way of keeping track 
of the Wick contractions, with a line for each Wick contraction and vertices 
for the $X$-matrices. The double line notation is more common in physics
for the large $N$ expansion, because the closed lines of the double line 
diagram give power of $N$, and when the double lines are thinned down 
to give the ribbon graph, the faces give the power of $N$. However 
as a way to keep track of the Wick contractions, it is clear that the 
transformed diagram is equally good and as shown by the argument above, 
this diagram of Wick contractions lives on the same genus Riemann surface. 
The transformation is similar to the one which was used 
in exhibiting the connection between QED Feynman graphs and ribbon graphs.

\end{appendix}

\end{document}